\documentclass[11pt]{article}
\usepackage[a4paper,margin=0.8in]{geometry}
\usepackage{graphicx}
\usepackage{physics}
\usepackage{derivative}
\usepackage{amsmath}
\usepackage{framed}
\usepackage{amssymb}
\usepackage{tikz}
\usepackage{booktabs} 
\usetikzlibrary{shapes.geometric, arrows.meta, positioning}
\usepackage[backend=bibtex, style=numeric-comp, sorting=none]{biblatex}
\DeclareUnicodeCharacter{211D}{\ensuremath{\mathbb{R}}}

\begin{document}
	
	\title{A Geometrically Parameterized Quasi-Stationary 3D Model for High-Frequency Induction Tube Welding}
	
	\author{
		D. Ivanov, J. I. Asperheim, B. Grande and P. Das
		\\
		\\
		ENRX a.s., Bølevegen 10, Skien, Norway~
	}	
	
	\date{}
	
	\maketitle
	
	\begin{abstract}
	A three-dimensional multiphysics finite element model for the simulation of high-frequency induction welding of tubes is presented. The model couples a time-harmonic $h-\phi$ formulation with a stabilized quasi-stationary advection–diffusion heat transport equation, enabling accurate prediction of electromagnetic and thermal fields under industrial operating conditions. The model incorporates parameterized geometry generation and semi-automated, physics-tailored mesh construction and is implemented using the open-source tools GetDP and Gmsh. Validation against measurements from a commercial induction welding line for AISI 304 stainless steel tubes demonstrates good agreement with operating data. The validated model is subsequently applied to investigate the influence of impeder material by comparing a conventional FeNiZnV ferrite with the soft magnetic composite Ferrotron 559H for the induction welding of AISI 304 stainless steel tubes.
	\end{abstract}
	
	\section{Introduction}
	
	High-frequency induction welding is one of the most widely used manufacturing processes for the production of longitudinally welded tubes. During the process, a continuously formed steel strip is heated by induced eddy currents and subsequently joined under pressure to produce a longitudinal weld. The underlying physics of the process is inherently complex, characterized by intricate electromagnetic field configurations, multiple current paths, and highly non-linear material behavior. Consequently, numerous numerical models have been developed to analyze this process \cite{asperheim1998temperature, asperheim2000temperature, asperheim2024numerical, das2020three, nikanorov2015numerical, muyskens2020improving, egger2022numerical}.
	
	Recently, we presented a comprehensive three-dimensional mathematical model for high-frequency induction welding that couples the time-harmonic Maxwell equations with a quasi-stationary advection--diffusion heat equation. The model incorporates realistic parameterized tube geometries with arbitrary Vee opening angles, spring-back effects, and spatially varying feed velocity, while employing stabilized finite element techniques to enable simulations at industrial welding-line speeds \cite{asperheim2024numerical, das2020three}. The formulation used enabled accurate three-dimensional simulations of the welding process, but it did not provide a natural way to prescribe global electrical circuit quantities, such as the coil voltage or current. Consequently, important characteristics of the induction system, including its electrical impedance, could not be evaluated directly. The magnetic vector potential formulation of Maxwell's equations was chosen. Although this formulation is well established and widely used, its numerical implementation is computationally demanding, particularly for large three-dimensional industrial models. This computational cost can be significantly reduced by employing alternative formulations, such as the $h-\phi$ formulation \cite{dular2019finite}.
	
	The present work targets further improvements to the simulation model, focusing on an alternative formulation that incorporates global circuit parameters, as well as a robust approach to geometry definition and parametrization. 
	
	The developed model is further validated using operational data from an existing high-frequency induction welding installation processing austenitic stainless steel AISI304 tubes. A major motivation for choosing this material is that its electro-thermo-physical properties are  well-established across the entire temperature range typical for tube welding process. Additionally, this non-magnetic material has a relatively high electrical resistivity, resulting in a large electromagnetic penetration depth. This reduces localized current density gradients, thereby lowering the mesh resolution requirements and easing the overall computational demand.
	
	The validated model is further used to address one of the key parameters of high-frequency induction welding, specifically investigating the influence of two distinct magnetic materials used for impeder: a soft magnetic composite (SMC) and a classical ferrite.
	
	\section{Mathematical Formulation}
	
	\subsection{Problem Domain and Geometric Decomposition}
	
	Let $\Omega \subset \mathbb{R}^3$ be a bounded computational domain partitioned into a conducting region $\Omega_c$ and a non-conducting complement domain $\Omega_c^C$, such that:
	\begin{equation}
		\Omega = \Omega_c \cup \Omega_c^C \quad \text{and} \quad \Omega_c \cap \Omega_c^C = \emptyset
	\end{equation}
	
	\begin{figure} [!h]
		\centering
		\includegraphics[width=0.6\linewidth]{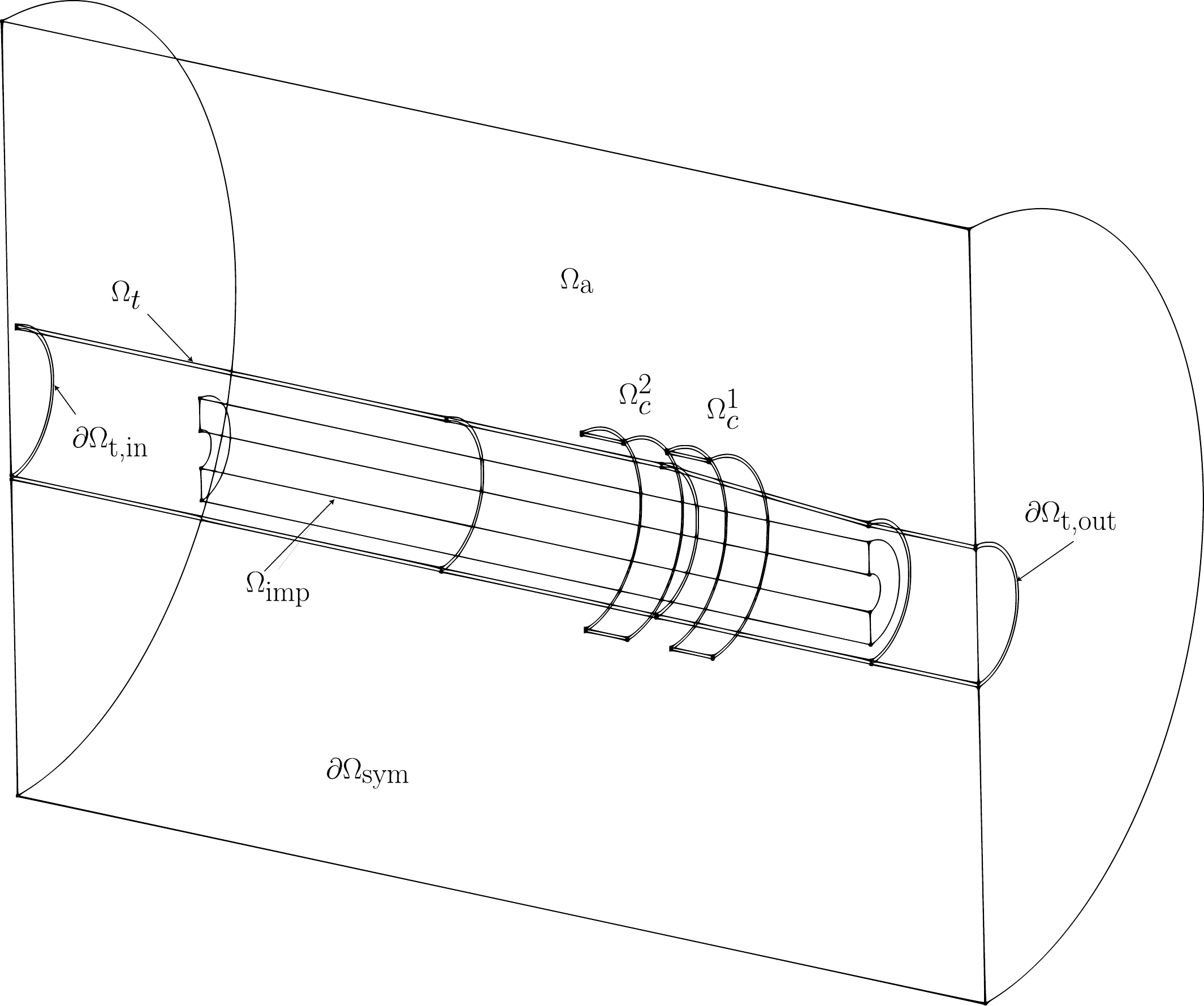}
		\caption{Simulation domains representing the welded tube, induction coil, impeder, and air}
		\label{fig:domain}
	\end{figure}
	
	The conducting region $\Omega_c$ is further subdivided into the welded tube strip and the multi-turn induction coil (see Fig.~\ref{fig:domain}):
	\begin{equation}
		\Omega_c = \Omega_t \cup \left( \bigcup_{i=1}^{N_c} \Omega_{c}^i \right)
	\end{equation}
	where $\Omega_t$ denotes the domain of the welded steel tube, and $\Omega_{c}^i$ represents the individual conducting volume of the $i$-th turn of the multi-turn induction coil ($i = 1, \dots, N_c$). Each coil turn carries a predefined total alternating current $I_i=I$ associated with an operating frequency $f$.
	
	The non-conducting surrounding medium $\Omega_c^C$ comprises the regions where macroscopic eddy currents are physically suppressed:
	\begin{equation}
		\Omega_c^C = \Omega_a \cup \Omega_{\text{imd}}
	\end{equation}
	where $\Omega_a$ represents the surrounding air domain, and $\Omega_{\text{imd}}$ isolates the magnetic impeder core positioned inside the tube (see Fig.~\ref{fig:domain}).
	
	The boundary of any subdomain $\Omega_x$ is denoted by the prefix $\partial$ followed by its corresponding symbol. In addition, several problem-specific boundaries are defined based on the geometric truncation of the setup. Due to the geometric and physical symmetry of the induction tube welding configuration, the computational domain is truncated using a symmetry boundary plane, denoted by $\partial\Omega_{\text{sym}}$. Furthermore, the truncated inlet and outlet boundaries of the tube domain ($\Omega_t$) are defined as $\partial\Omega_{t,\text{in}}$ and $\partial\Omega_{t,\text{out}}$, respectively.
	
	\subsection{Electromagnetic Model}
	
	\subsubsection{Governing Maxwell Equations}
	
	The magnetodynamic problem is governed by the eddy current approximation of Maxwell's equations, where the displacement current density $\partial_t \mathbf{d}$ is neglected \cite{geuzaine2001high}:
	\begin{align}
		\mathbf{curl} \, \mathbf{h} &= \mathbf{j} && \text{in } \Omega \quad \text{(Ampère's Law)} \label{eq:ampere} \\
		\mathbf{curl} \, \mathbf{e} &= -\partial_t \mathbf{b} && \text{in } \Omega \quad \text{(Faraday's Law)} \label{eq:faraday} \\
		\mathbf{div} \, \mathbf{b} &= 0 && \text{in } \Omega \quad \text{(Gauss's Law)} \label{eq:gauss_mag}
	\end{align}
	Here, $\mathbf{h}$ is the magnetic field intensity, $\mathbf{b}$ is the magnetic flux density, $\mathbf{e}$ is the electric field intensity, and $\mathbf{j}$ is the electric current density.
	
	The system is closed by the non-linear, temperature-dependent constitutive laws representing the material properties:
	\begin{align}
		\mathbf{b} &= \mu(|\mathbf{h}|, \theta) \, \mathbf{h} \label{eq:mu} \\
		\mathbf{e} &= \rho(\theta) \, \mathbf{j} \label{eq:rho}
	\end{align}
	Here, $\mu$ is the magnetic permeability, $\rho(\theta)$ is the temperature-dependent electrical resistivity ($\rho > 0$ inside $\Omega_c$, and $\rho \to \infty$ in $\Omega_c^C$), and $\theta$ is the localized temperature field.
	
	Substituting Ampère's law \eqref{eq:ampere} into Ohm's law \eqref{eq:rho}, and inserting the resulting expression for the electric field $\mathbf{e}$ into Faraday's law of induction \eqref{eq:faraday} while incorporating the non-linear magnetic constitutive relation \eqref{eq:mu}, yields the single, unified partial differential equation governing the magnetic field intensity $\mathbf{h}$ across the entire computational space $\Omega$:
	\begin{equation}
		\mathbf{curl} \, \big(\rho(\theta) \, \mathbf{curl} \, \mathbf{h}\big) + \partial_t \big(\mu(|\mathbf{h}|, \theta) \, \mathbf{h}\big) = \mathbf{0} \quad \text{in } \Omega \label{eq:pde_cond}
	\end{equation}
	
	In the non-conducting domain $\Omega_c^C$, the electrical resistivity approaches infinity ($\rho \to \infty$), meaning no electric current density can exist ($\mathbf{j} = \mathbf{0}$). Consequently, the fields there must satisfy a strict irrotational constraint:
	\begin{align}
		\mathbf{curl} \, \mathbf{h} &= \mathbf{0} \quad \text{in } \Omega_c^C \label{eq:pde_air_curl}
	\end{align}
	
	\subsubsection{Field Decomposition, Topology of the Domain $\Omega$, and Total Coil Current $I$}
	\label{sssec:cuts}
\begin{figure} [!h]
		\centering
		\includegraphics[width=0.6\linewidth]{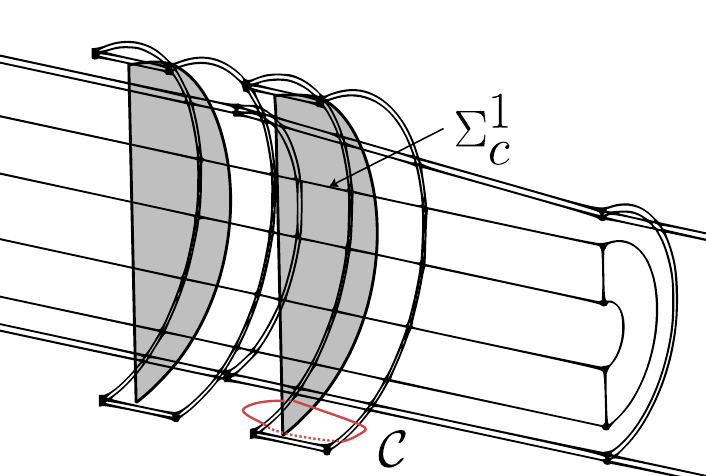}
		\caption{Domain cuts essential for handling the multi-connected topology of the insulator domain.}
		\label{fig:cuts}
	\end{figure}

	The magnetic field intensity $\mathbf{h}$ in the non-conducting region $\Omega_c^C$ is governed by the absence of electric currents \eqref{eq:pde_air_curl}. While this property theoretically allows $\mathbf{h}$ to be represented solely by the gradient of a total, single-valued magnetic scalar potential $\phi$ via $\mathbf{h} = -\mathbf{grad} \, \phi$, a topological conflict arises if the insulator subdomain $\Omega_c^C$ is multi-connected (i.e., it contains topological tunnels through which the currents flow) \cite{pellikka2013homology}. In the case of $\Omega_c^C$, the turns of an induction coil render the surrounding air domain multi-connected. In such configurations, one can define a closed contour $\mathcal{C}$ (see Fig.~\ref{fig:cuts}) within $\Omega_c^C$ that links a current loop; according to the integral form of Ampère's Law, this contour must satisfy:
	\begin{equation}
		\oint_{\mathcal{C}} \mathbf{h} \cdot \text{d}\mathbf{l} = I \neq 0 \label{eq:amperes_law}
	\end{equation}
	
	Because the line integral of a gradient field along a closed loop is identically zero ($\oint \mathbf{grad} \, \phi \cdot \text{d}\mathbf{l} = 0$), a continuous scalar potential cannot account for this net enclosed current. 
	
	In the standard approach, the topological conflict is resolved by introducing artificial geometric cuts $\Sigma_c^i$ (see Fig.~\ref{fig:cuts}) for each independent current loop (induction coil turn $i$ carrying current $I$) to restore a simply-connected topology to the insulator subdomain $\Omega_c^C$ \cite{pellikka2013homology, geuzaine2001high, dlotko2010efficient}. Mathematically, this is achieved by introducing current basis functions $\mathbf{c}_i$ (associated with the cuts). These functions are defined as the gradient of a local potential $\phi_{d,i}$ that exhibits a unit step-jump across the cut surface $\Sigma_c^i$ \cite{geuzaine2001high}:
	\begin{equation}
		\mathbf{c}_i = -\mathbf{grad} \, \phi_{d,i} \quad \text{with} \quad [\phi_{d,i}]_{\Sigma_c^i} = 1 \label{eq:cut_basis}
	\end{equation}
	Consequently, the full discrete decomposition of the magnetic field intensity $\mathbf{h}$ across the entire computational space $\Omega$ is expressed as:
	\begin{equation}
		\mathbf{h} = -\mathbf{grad} \, \phi + \mathbf{t} + \sum_{i=1}^{N_c} I_i \mathbf{c}_i \label{eq:h_full_decomposition}
	\end{equation}
	where $\mathbf{t}$ represents the field restricted to the conducting domain $\Omega_c$ to support local eddy currents, and $I_i$ acts as the macroscopic net current coefficient associated with the topological cuts.
	
	\subsubsection{Interface Continuity}
	
	At the internal material interface $\partial\Omega_c \cap \partial\Omega_c^C$ separating the conducting domain $\Omega_c$ from the surrounding air $\Omega_c^C$, the fields must respect fundamental physical transition laws. Because the bulk electrical conductivity remains finite within the welded tube and induction coil, no mathematical surface current densities exist at the boundary. Consequently, the tangential component of the magnetic field intensity must be strictly continuous across the interface. Furthermore, Gauss's law for magnetism \eqref{eq:gauss_mag} dictates that the magnetic flux density must remain globally solenoidal, which strictly implies the continuity of its normal component across any material discontinuity. These interface conditions are enforced by:
	\begin{align}
		\mathbf{n} \times (\mathbf{h}_c - \mathbf{h}_a) &= \mathbf{0} \quad \text{on } \partial\Omega_c \cap \partial\Omega_c^C \label{eq:cont_tangential_h} \\
		\mathbf{n} \cdot (\mathbf{b}_c - \mathbf{b}_a) &= 0 \quad \text{on } \partial\Omega_c \cap \partial\Omega_c^C  \label{eq:cont_normal_b}
	\end{align}
	where $\mathbf{n}$ is the unit normal vector to the boundary, while $\mathbf{h}_c, \mathbf{b}_c$ and $\mathbf{h}_a, \mathbf{b}_a$ represent the magnetic field limits approaching from the conductor and air sides, respectively.
	
	\subsubsection{Boundary Conditions}
	To yield a unique solution for the algebraic system, the computational domain must be truncated using appropriate boundary conditions on the far outer facets and symmetry plane. 
	
	Assuming the outer boundaries are placed sufficiently far from the induction coil and workpiece, the magnetic field lines wrap parallel to the boundary limits. This behavior is captured by enforcing a vanishing tangential $\mathbf{h}$-field component:
	\begin{equation}
		\mathbf{h} \times \mathbf{n} = \mathbf{0} \quad \text{on } \partial\Omega \setminus \partial\Omega_{\text{sym}} \label{eq:bc_magnetic_wall}
	\end{equation}
	
	Furthermore, to reduce computational cost, geometric and physical symmetries are exploited by applying mirror conditions on $\partial\Omega_{\text{sym}}$. We enforce a vanishing tangential magnetic field component on the symmetry plane:
	\begin{equation}
		\mathbf{h} \times \mathbf{n} = \mathbf{0} \quad \text{on } \partial\Omega_{\text{sym}} \label{eq:bc_mirror}
	\end{equation}
	
	\subsubsection{Time-Harmonic (Complex-Valued) Formulation}
	
	When the excitation current varies purely sinusoidally over time at an angular frequency $\omega = 2\pi f$, the transient governing equations can be efficiently cast into the frequency domain. Under this steady-state sinusoidal assumption, the real-valued, time-dependent magnetic field intensity $\mathbf{h}(t)$ is represented by the real part of its corresponding complex-valued phasor field $\underline{\mathbf{h}}$ scaling a harmonic time-factor:
	\begin{equation}
		\mathbf{h}(t) = \Re \big\{ \underline{\mathbf{h}} \, e^{j\omega t} \big\} \label{eq:phasor_definition}
	\end{equation}
	where $j = \sqrt{-1}$ is the imaginary unit. Applying this transformation to the Maxwell system converts the temporal derivative operator into an algebraic multiplication: $\partial_t \to j\omega$.
	
	Consequently, the unified partial differential equation \eqref{eq:pde_cond} governing the complex magnetic field intensity phasor $\underline{\mathbf{h}}$ yields the complex-valued time-harmonic formulation across $\Omega$:
	\begin{equation}
		\mathbf{curl} \, \big(\rho(\theta) \, \mathbf{curl} \, \underline{\mathbf{h}}\big) + j\omega \big(\mu_{\text{eff}}(|\underline{\mathbf{h}}|, \theta) \, \underline{\mathbf{h}}\big) = \mathbf{0} \quad \text{in } \Omega \label{eq:pde_harmonic}
	\end{equation}
	where $\mu_{\text{eff}}(|\underline{\mathbf{h}}|, \theta)$ is the precalculated effective magnetic permeability. 
	
	The need for introducing the mapping $\mu \rightarrow \mu_{\text{eff}}$ arises from the fact that, even in the case of a purely sinusoidal excitation, the non-linear magnetic characteristics of the impeder and ferromagnetic steel distort the harmonic nature of the fields within the magnetic regions once the peak field intensity exceeds the linear threshold. To preserve the computational advantages of the single-frequency complex phasor framework within these non-linear regimes, an effective permeability $\mu_{\text{eff}}$ is adopted. Following a classical co-energy conservation approach \cite{paoli2002complex}, this effective permeability is derived directly from the material's standard dehysterized $B\text{-}H$ characteristic curve:
	\begin{equation}
		\mu_{\text{eff}}(H) = \frac{2}{H^2} \int_{0}^{H} B(\hat{H}) \, \text{d}\hat{H} \label{eq:mu_effective_coenergy}
	\end{equation}
	It is important to note that the introduction of $\mu_{\text{eff}}$ for non-linear magnetic materials is not mathematically rigorous; it merely approximates the solution using a single fundamental frequency. In reality, the true physical response within these non-linear regions is inherently multi-harmonic.
	
	\subsection{Heat Transfer}
	
	\subsubsection{Governing Heat Transport Equation}
	
	The thermal problem within the moving workpiece is governed by a quasi-stationary advection-diffusion equation \cite{asperheim2024numerical, das2020three}. Let $\Omega_t \subset \Omega_c$ denote the conducting subdomain representing the steel tube undergoing geometric forming and continuous feeding through the welding section. The spatial and temporal distribution of the temperature field $\theta$ is described by:
	\begin{equation}
		\rho_m(\theta) c_p(\theta) \, \mathbf{v} \cdot \mathbf{grad} \, \theta - \mathbf{div} \, \big(k(\theta) \, \mathbf{grad} \, \theta\big) = Q_{\text{src}} \quad \text{in } \Omega_t \label{eq:heat_transport}
	\end{equation}
	where $c_p(\theta)$ is the temperature-dependent specific heat capacity, $\rho_m$ is the constant mass density of the material, and $k(\theta)$ is the temperature-dependent thermal conductivity. 
	
	The first term in \eqref{eq:heat_transport} accounts for the convective heat transport due to the material motion, where $\mathbf{v}$ denotes the prescriptive local velocity vector of the moving steel strip. The second term represents the heat transfer within the domain via thermal conduction driven by spatial temperature gradients.
	
	\subsubsection{Boundary Conditions}
	
	To yield a unique solution to the elliptic differential equation \eqref{eq:heat_transport}, appropriate thermal boundary conditions must be prescribed along the boundary surfaces of the tube domain $\partial\Omega_t$. 
	
	At the upstream inlet cross-section of the tube $\partial\Omega_{t,\text{in}}$, the incoming material is assumed to be at a known baseline temperature. This is enforced via a standard Dirichlet boundary condition:
	\begin{equation}
		\theta = \theta_{\text{in}} \quad \text{on } \partial\Omega_{t,\text{in}} \label{eq:thermal_dirichlet}
	\end{equation}
	where $\theta_{\text{in}}$ is the prescribed ambient entry temperature of the steel strip before entering the induction heating zone.
	
	To account for the complex heat dissipation across the tube surfaces exposed to the surrounding environment, a non-linear Robin (mixed) boundary condition is implemented. This boundary condition accounts for the convective heat losses to the ambient air:
	\begin{equation}
		-k(\theta) \, \mathbf{grad} \, \theta \cdot \mathbf{n} = h_{\text{conv}}(\theta - \theta_0)  \quad \text{on }  \partial\Omega_t \cap \partial\Omega_a \label{eq:Robin_condition}
	\end{equation}
	where $h_{\text{conv}}$ represents the convective heat transfer coefficient and $\theta_0$ is the air temperature.
	
	\subsection{Electromagnetic–Thermal Coupling}
	
	The volumetric heat source term $Q_{\text{src}}$ drives the thermal transport equation \eqref{eq:heat_transport} and represents the localized macroscopic energy dissipation. In modeling the high-frequency induction tube welding process, the following physical assumptions are established to define the heat source distribution:
	\begin{enumerate}
		\item \textbf{Neglection of Non-Ohmic Losses:} Power dissipation due to magnetic hysteresis cycles is neglected within the tube domain ($\Omega_t$), as the process is dominated by eddy-current driven Joule heating. Furthermore, internal energy dissipation and localized heat generation resulting from mechanical deformation during the tube-forming (\cite{egger2022numerical}) are assumed to be negligible compared to the high-intensity electromagnetic power input.
		\item \textbf{Separation of Time Scales:} The macroscopic thermal time constant of the moving tube is orders of magnitude larger than the microsecond electromagnetic period $T = 1/f$. This wide separation allows the multi-physical fields to be decoupled by utilizing a steady, time-averaged power density.
	\end{enumerate}
	
	By adopting the frequency-domain framework, this time-averaged power density is computed directly from the complex-valued phasor fields without temporal integration. Given the peak value phasor of the current density $\underline{\mathbf{j}}$, the mean volumetric heat source is expressed exclusively as:
	\begin{equation}
		Q_{\text{src}} = \frac{1}{2} \rho(\theta) \, \underline{\mathbf{j}} \cdot \underline{\mathbf{j}}^* = \frac{1}{2} \rho(\theta) \, |\underline{\mathbf{j}}|^2 \label{eq:joule_heating_complex}
	\end{equation}
	where $\underline{\mathbf{j}}^*$ denotes the complex conjugate of the current density phasor field, which is linked to the complex magnetic field intensity phasor via $\underline{\mathbf{j}} = \mathbf{curl} \, \underline{\mathbf{h}}$.
	
	\section{Numerical Implementation}
	
	\subsection{Electromagnetic Problem}
	
	\subsubsection{Weak Formulation}
	We seek a weak solution $\underline{\mathbf{h}}$ of the governing equation \eqref{eq:pde_harmonic} that lives in the curl-conforming $H(\text{curl}; \Omega)$ space \cite{monk2003finite}:
	\begin{equation}
		H(\text{curl}; \Omega) = \left\{ \mathbf{v} \in L^2(\Omega)^3 \;\middle|\; \mathbf{curl} \, \mathbf{v} \in L^2(\Omega)^3 \right\}
	\end{equation}
	
	It is important to note that $H(\text{curl}; \Omega)$ is a larger, functionally "weaker" space than the standard vector Sobolev space $H^1(\Omega)^3$, meaning that $H^1(\Omega)^3 \subset H(\text{curl}; \Omega)$. While any function in $H^1$ must possess square-integrable weak derivatives for all its components in all directions, a function in $H(\text{curl})$ is only required to have a well-defined weak curl which is consistent with continuity of the tangential field across interfaces, as expressed in \eqref{eq:cont_tangential_h}. At the same time, this space does not enforce continuity of the normal component of $\underline{\mathbf{h}}$, allowing jumps in the normal field at
	material interfaces. This behavior aligns with
	\eqref{eq:cont_normal_b}.
		
	To satisfy the outer boundary conditions \eqref{eq:bc_magnetic_wall} and symmetry conditions \eqref{eq:bc_mirror} in a strong sense, we restrict our attention to the constrained Hilbert space $H_{\mathbf{h}}^0(\text{curl}; \Omega) \subset H(\text{curl}; \Omega)$, defined as:
	\begin{equation}
		H_{\mathbf{h}}^0(\text{curl}; \Omega) = \left\{ \mathbf{v} \in H(\text{curl}; \Omega) \;\middle|\; \mathbf{v} \times \mathbf{n} = \mathbf{0} \text{ on } \partial\Omega \right\}
	\end{equation}
	
	Following the standard Galerkin approach, we multiply the governing equation \eqref{eq:pde_harmonic} by a time-independent complex conjugate test function $\underline{\mathbf{h}}' \in H_{\mathbf{h}}^0(\text{curl}; \Omega)$ chosen from the same functional space. Integrating the resulting expression over the global domain $\Omega$ and integrating by parts yields:
	
	\begin{equation}
		\int_{\Omega} j\omega \, \mu_{\text{eff}}(|\underline{\mathbf{h}}|, \theta) \underline{\mathbf{h}} \cdot \underline{\mathbf{h}}' \, \text{d}\Omega + \int_{\Omega} \rho(\theta) (\mathbf{curl} \, \underline{\mathbf{h}}) \cdot (\mathbf{curl} \, \underline{\mathbf{h}}') \, \text{d}\Omega + \int_{\partial\Omega} (\mathbf{n} \times \mathbf{e}) \cdot \underline{\mathbf{h}}' \, \text{d}\Gamma = 0
	\end{equation}
	Because the test functions satisfy $\underline{\mathbf{h}}' \times \mathbf{n} = \mathbf{0}$ on the boundary $\partial\Omega$, the surface integral vanishes. Furthermore, utilizing the definition of the non-conducting domain $\Omega_c^C$ where $\rho(\theta) \to \infty$ and $\mathbf{curl} \, \underline{\mathbf{h}} = \mathbf{0}$ via \eqref{eq:pde_air_curl}, the continuous weak variational statement is formulated as follows:
	
	Find the continuous trial phasor field $\underline{\mathbf{h}} \in H_{\mathbf{h}}^0(\text{curl}; \Omega)$ such that:
	\begin{equation}
		\int_{\Omega} j\omega \, \mu_{\text{eff}}(|\underline{\mathbf{h}}|, \theta) \underline{\mathbf{h}} \cdot \underline{\mathbf{h}}' \, \text{d}\Omega + \int_{\Omega_c} \rho(\theta) \left( \mathbf{curl} \, \underline{\mathbf{h}} \right) \cdot \left( \mathbf{curl} \, \underline{\mathbf{h}}' \right) \, \text{d}\Omega = \mathbf{0}, \quad \forall \underline{\mathbf{h}}' \in H_{\mathbf{h}}^0(\text{curl}; \Omega) \label{eq:continuous_weak_statement}
	\end{equation}
	
	\subsubsection{Discretization}
	\label{sssec:elmag_discret}
	
	The continuous weak formulation \eqref{eq:continuous_weak_statement} is discretized using conforming finite-dimensional spaces. Following the local--global field decomposition framework of Dular et al.~\cite{dular1998coupling}, the discrete field $\underline{\mathbf{h}}_h$ is approximated using curl-conforming Nédélec basis functions (equivalently, Whitney 1-forms) \cite{nedelec1980mixed}, together with compatible nodal contributions for the scalar potential representation:
	
	\begin{equation}
		\underline{\mathbf{h}}_h = -\mathbf{grad} \left( \sum_{n \in \mathcal{N}} \underline{\phi}_n \alpha_n \right) + \sum_{e \in \mathcal{E}} \underline{t}_e \mathbf{s}_e + \sum_{i=1}^{N_c} \underline{I}_i \mathbf{c}_i , \label{eq:h_discrete_decomposition}
	\end{equation}
	where $\underline{\phi}_n$ and $\underline{t}_e$ represent the unknown complex nodal and edge degrees of freedom, and $\underline{I}_i$ is the prescribed current in the $i$-th coil turn. 
	
	The conformity of each component in \eqref{eq:h_discrete_decomposition} with $H(\mathrm{curl};\Omega)$ follows directly from the construction of the underlying basis functions:
	\begin{itemize}
		\item $\alpha_n$ are Whitney 0-forms (continuous nodal functions in $H^1(\Omega)$). Because $\mathbf{curl} \, \mathbf{grad} \, \alpha_n = \mathbf{0}$, the term $-\mathbf{grad}\,\alpha_n$ is trivially curl-conforming.
		
		\item $\mathbf{s}_e$ are Whitney 1-forms (edge basis functions), which are curl-conforming by construction. Restricting the active edge set to interior edges of the conducting region ($\mathcal E \cap \partial\Omega_c = \varnothing$) strongly enforces the homogeneous tangential interface condition on $\partial\Omega_c$.
		
		\item $\mathbf{c}_i \in \mathcal{W}^1(\Omega)$ are global cohomology basis functions constructed as linear combinations of Whitney 1-forms \cite{dular1998coupling}, ensuring tangential continuity while providing the non-trivial circulation required to represent the current linkage prescribed by Amp\`ere's law.
	\end{itemize}
	
	Following the standard Galerkin method, the test functions $\underline{\mathbf{h}}'$ are selected from the same approximation space. For current-driven excitation, the coil currents $\underline{I}_i$ are prescribed, meaning their virtual variations vanish ($\underline{I}_i' = 0$). The test functions are thus generated solely by the local scalar and edge variations:
	\begin{equation}
		\underline{\mathbf{h}}' = -\mathbf{grad} \left( \sum_{m \in \mathcal{N}} \underline{\phi}'_m \alpha_m \right) + \sum_{k \in \mathcal{E}} \underline{t}'_k \mathbf{s}_k . \label{eq:h_test_discrete}
	\end{equation}
	
	Substituting \eqref{eq:h_discrete_decomposition} and \eqref{eq:h_test_discrete} into the continuous weak formulation yields a
	natural algebraic separation between the unknown field coefficients and the prescribed current excitations:
	\begin{align}
		&\int_{\Omega} j\omega \, \mu_{\text{eff}}(|\underline{\mathbf{h}}_h|,\theta) \left( -\mathbf{grad} \sum_{n \in \mathcal N} \underline{\phi}_n \alpha_n + \sum_{e \in \mathcal E} \underline t_e \mathbf s_e \right) \cdot \underline{\mathbf h}' \, \mathrm d\Omega + \int_{\Omega_c} \rho(\theta)\, \mathbf{curl} \left( \sum_{e \in \mathcal E} \underline t_e \mathbf s_e \right) \cdot \mathbf{curl}\, \underline{\mathbf h}' \, \mathrm d\Omega \nonumber\\
		&\quad = - \sum_{i=1}^{N_c} \underline I_i \left( \int_{\Omega} j\omega\, \mu_{\text{eff}}(|\underline{\mathbf h}_h|,\theta) \,\mathbf c_i \cdot \underline{\mathbf h}' \,\mathrm d\Omega + \int_{\Omega_c} \rho(\theta)\, (\mathbf{curl}\,\mathbf c_i) \cdot (\mathbf{curl}\,\underline{\mathbf h}') \,\mathrm d\Omega \right). \label{eq:weak_separated}
	\end{align}

	\subsubsection{Post-Processing of Terminal Voltages}

	Since the system is driven by prescribed current sources, the terminal voltage phasors $\underline{V}_i$ are not primary unknowns in \eqref{eq:weak_separated}. They are recovered in a post-processing step from the computed magnetic field $\underline{\mathbf{h}}_h$.
	
	In a classical formulation, $\underline{V}_i$ would be evaluated as a cut-surface integral associated with the $i$-th coil loop. In the present approach, the cut is represented implicitly through the cohomology basis functions $\mathbf{c}_i \in \mathcal{W}^1(\Omega)$.
	
	The terminal voltage is obtained via the duality pairing
	\begin{equation}
		\underline{V}_i =
		\int_{\Omega}
		j\omega \, \mu_{\text{eff}}(|\underline{\mathbf{h}}_h|,\theta)
		\, \underline{\mathbf{h}}_h \cdot \mathbf{c}_i \, \mathrm{d}\Omega
		+
		\int_{\Omega_c}
		\rho(\theta)
		\left( \mathbf{curl} \, \underline{\mathbf{h}}_h \right)
		\cdot
		\left( \mathbf{curl} \, \mathbf{c}_i \right)
		\, \mathrm{d}\Omega .
		\label{eq:voltage}
	\end{equation}
	
	\subsection{Heat Transfer Problem}
	
	\subsubsection{Weak Formulation}
	We seek a weak solution $\theta$ of the governing thermal transport equation that lives in the standard Sobolev space $H^1(\Omega_t)$ \cite{brenner2008mathematical}:
	\begin{equation}
		H^1(\Omega_t) = \left\{ v \in L^2(\Omega_t) \;\middle|\; \mathbf{grad} \, v \in L^2(\Omega_t)^3 \right\}
	\end{equation}
		
	To strongly enforce the non-homogeneous Dirichlet boundary condition \eqref{eq:thermal_dirichlet} at the upstream entry cross-section, we define the constrained trial space $H^1_{\theta_{\text{in}}}(\Omega_t) \subset H^1(\Omega_t)$ and the corresponding homogeneous test space $H^1_0(\Omega_t) \subset H^1(\Omega_t)$ as:
	\begin{align}
		H^1_{\theta_{\text{in}}}(\Omega_t) &= \left\{ v \in H^1(\Omega_t) \;\middle|\; v = \theta_{\text{in}} \text{ on } \partial\Omega_{t,\text{in}} \right\} \\
		H^1_0(\Omega_t) &= \left\{ v \in H^1(\Omega_t) \;\middle|\; v = 0 \text{ on } \partial\Omega_{t,\text{in}} \right\}
	\end{align}
	
	Following the standard Galerkin approach, we multiply the governing equation \eqref{eq:heat_transport} by a time-independent scalar test function $\theta' \in H^1_0(\Omega_t)$. Integrating over the computational domain $\Omega_t$ and applying integration by parts to the diffusion term yields
	\begin{align}
		\int_{\Omega_t} \rho_m(\theta) c_p(\theta) \left( \mathbf{v} \cdot \mathbf{grad} \, \theta \right) \theta' \, \text{d}\Omega &+ \int_{\Omega_t} k(\theta) \, \mathbf{grad} \, \theta \cdot \mathbf{grad} \, \theta' \, \text{d}\Omega \nonumber \\
		&- \int_{\partial\Omega_t} \big(k(\theta) \, \mathbf{grad} \, \theta \cdot \mathbf{n}\big) \theta' \, \text{d}\Gamma = \int_{\Omega_t} Q_{\text{src}} \, \theta' \, \text{d}\Omega
	\end{align}
	
	The boundary integral vanishes identically on the inlet $\partial\Omega_{t,\text{in}}$ because $\theta' = 0$, and on the boundaries $\partial\Omega_{t,\text{sym}}$ and $\partial\Omega_{t,\text{out}}$ due to the homogeneous Neumann (adiabatic) condition. Substituting the non-linear convective Robin condition \eqref{eq:Robin_condition} over the exposed external surface area $\partial\Omega_t \cap \partial\Omega_a$, the continuous weak variational statement is formulated as follows:
	
	Find the continuous temperature field $\theta \in H^1_{\theta_{\text{in}}}(\Omega_t)$ such that:
	\begin{align}
		\int_{\Omega_t} \rho_m(\theta) c_p(\theta) \left( \mathbf{v} \cdot \mathbf{grad} \, \theta \right) \theta' \, \text{d}\Omega 
		&+ \int_{\Omega_t} k(\theta) \, \mathbf{grad} \, \theta \cdot \mathbf{grad} \, \theta' \, \text{d}\Omega \nonumber \\
		&+ \int_{\partial\Omega_t \cap \partial\Omega_a}  h_{\text{conv}}(\theta - \theta_0) \theta' \, \text{d}\Gamma 
		= \int_{\Omega_t} Q_{\text{src}} \, \theta' \, \text{d}\Omega, \quad \forall \theta' \in H^1_0(\Omega_t) \label{eq:thermal_weak_statement}
	\end{align}
	
	\subsubsection{Stabilized Discrete Formulation}
	\label{sssec:thermal_discret}
	
	The continuous weak formulation \eqref{eq:thermal_weak_statement} is discretized using conforming finite-dimensional spaces. Following the standard continuous Galerkin approach, the discrete temperature field $\theta_h$ is approximated using first-order nodal Lagrange basis functions (equivalently, Whitney 0-forms):
	\begin{equation}
		\theta_h = \sum_{n \in \mathcal{N}_0} \theta_n \alpha_n + \sum_{b \in \mathcal{N}_{\text{in}}} \theta_{\text{in}, b} \alpha_b , \label{eq:theta_discrete_decomposition}
	\end{equation}
	where $\theta_n$ denotes the unknown scalar degree of freedom associated with node $n$; $\theta_{\text{in}, b}$ is the prescribed inlet temperature at node $b$, while $\mathcal{N}_0$ and $\mathcal{N}_{\text{in}}$ denote the active internal and Dirichlet boundary node subsets, respectively ($\mathcal{N} = \mathcal{N}_0 \cup \mathcal{N}_{\text{in}}$).
	
	To preserve numerical stability under dominant advection regimes, the Streamline Upwind Petrov-Galerkin (SUPG) method is employed. While the trial field remains within the space spanned by \eqref{eq:theta_discrete_decomposition}, the non-symmetric test functions $\psi_m$ associated with each active node $m \in \mathcal{N}_0$ are systematically modified element-by-element along the velocity streamlines:
	\begin{equation}
		\psi_m = \alpha_m + \tau_e (\mathbf{v} \cdot \mathbf{grad} \, \alpha_m) , \label{eq:theta_test_discrete}
	\end{equation}
	where $\tau_e$ represents the element-dependent stabilization parameter \cite{john2011posteriori}.
	
	Following the unified framework for optimal residual-based stabilization parameters \cite{john2011posteriori}, the local element-wise Peclet number $Pe$ and the choice of $\tau_e$ are evaluated on each element of the domain $\Omega_t$ as follows:
	\begin{equation}
		Pe = \frac{\|\mathbf{v}\| \, h_e}{2 \lambda(\theta)}, \quad \tau_e = \frac{h_e}{2 \|\mathbf{v}\|} \left( \coth(Pe) - \frac{1}{Pe} \right) \label{eq:peclet_definition}
	\end{equation}
	
	Substituting the discrete trial \eqref{eq:theta_discrete_decomposition} and test functions \eqref{eq:theta_test_discrete} into the continuous weak formulation \eqref{eq:thermal_weak_statement} yields the stabilized discrete variational statement:
	\begin{align}
		&\int_{\Omega_t} k(\theta_h) \, \mathbf{grad} \, \theta_h \cdot \mathbf{grad} \, \alpha_m \, \text{d}\Omega \nonumber \\
		&+ \int_{\Omega_t} \rho_m(\theta_h) c_p(\theta_h) \left( \mathbf{v} \cdot \mathbf{grad} \, \theta_h \right) \alpha_m \, \text{d}\Omega \nonumber \\
		&+ \sum_{e} \int_{\Omega_e} \tau_e \rho_m(\theta_h) c_p(\theta_h) \left( \mathbf{v} \cdot \mathbf{grad} \, \theta_h \right) (\mathbf{v} \cdot \mathbf{grad} \, \alpha_m) \, \text{d}\Omega \nonumber \\
		&- \int_{\Omega_t} Q_{\text{src}} \, \alpha_m \, \text{d}\Omega - \sum_{e} \int_{\Omega_e} \tau_e Q_{\text{src}} (\mathbf{v} \cdot \mathbf{grad} \, \alpha_m) \, \text{d}\Omega \nonumber \\
		&+ \int_{\partial\Omega_t \cap \partial\Omega_a} h_{\text{conv}}(\theta_h - \theta_0) \alpha_m \, \text{d}\Gamma = 0 .
		\label{eq:thermal_discrete}
	\end{align}
	This formulation assumes linear finite elements, where the temperature gradient is constant within each element. Consequently, the diffusion contribution to the strong residual in the Streamline-Upwind/Petrov-Galerkin (SUPG) stabilization is omitted, matching the elementwise vanishing Laplacian inherent to linear elements under locally constant thermal conductivity.
	
	\subsection{Numerical Implementation within the GetDP Solver Framework}
	
	The coupled electromagnetic--thermal model is implemented using the open-source finite element solver GetDP \cite{dular1998general}. GetDP provides a flexible framework for the formulation of multiphysics problems through user-defined weak variational statements and a broad collection of conforming finite element function spaces. In addition to standard finite element discretizations, it supports the definition of global quantities and circuital constraints, which are particularly useful for the cohomology-based electromagnetic formulation employed in this work \cite{dular1998coupling, geuzaine2001high}. The solver is tightly integrated with Gmsh \cite{geuzaine2009gmsh}, which is used for geometry definition and mesh generation. For multiply connected domains, Gmsh can also automatically compute the cohomology basis functions (cuts) required by the local--global field decomposition.

	The general implementation of numerical models in GetDP follows a standard declarative workflow structured around the sequential definition of geometric groups, function spaces, formulations, and resolutions \cite{dular1998general, getdp_online}. 
	
	GetDP solver lacks a native mechanism to compute the element-dependent stabilization parameter $\tau_e$ for non-symmetric Petrov-Galerkin test spaces. Consequently, implementing the stabilized thermal formulation \eqref{eq:thermal_discrete} required low-level modifications of the solver's core engine. We implemented a new low-level runtime function, \texttt{TimeScale\_SUPG[]}, which operates strictly on the element level to dynamically evaluate the localized grid Peclet number $Pe$ and compute the element-dependent stabilization parameter $\tau_e$ for each element interior via \eqref{eq:peclet_definition}. This tailored implementation is integrated into an open-source fork of the solver, \texttt{ENRXGetDP} \cite{enrxgetdp2026}.

	\subsubsection{Coupling Scheme and Resolution Strategy}

	In the proposed FEM model, the excitation current $I$ and the operating frequency $f$ are prescribed input parameters used to compute the electromagnetic field and the resulting temperature distribution within the welded tube. In industrial practice, however, the inverse problem is often of greater interest: determining the electrical operating conditions required to achieve a prescribed thermal state. In particular, the objective is to determine the coil current required to reach a specified temperature at the squeeze-roll weld point (V-apex). To this end, an outer root-finding procedure is introduced to determine the current $I_{\mathrm{target}}$ for a prescribed frequency $f$ such that the weld-point temperature matches the desired value $\theta_{\mathrm{target}}$.

	Let $m$ denote the outer iteration index. Convergence is achieved when the temperature error at the weld point satisfies
	\begin{equation}
		\left| \theta_{\mathrm{weld}}^{m} - \theta_{\mathrm{target}} \right|
		\le
		\epsilon_{\mathrm{target}},
	\end{equation}
	where $\epsilon_{\mathrm{target}}$ is the prescribed temperature tolerance.

	To solve the nonlinear equation
	\[
	\theta_{\mathrm{weld}}(I)-\theta_{\mathrm{target}}=0,
	\]
	a modified secant method is implemented. At the initial iteration ($m=0$), the second current estimate is obtained using a physically motivated scaling based on the quadratic dependence of Joule heating on the excitation current ($Q_{\mathrm{src}}\propto I^2$). Assuming that the temperature rise scales approximately with the generated heat, the initial update is given by
	\begin{equation}
		I^{m+1}
		=
		I^{m}
		\sqrt{
			\frac{\theta_{\mathrm{target}}}
			{\theta_{\mathrm{weld}}^{m}}
		}.
	\end{equation}
	
	For subsequent iterations ($m\ge1$), the current is updated using the standard secant method,
	\begin{equation}
		I^{m+1}
		=
		I^{m}
		-
		\left(
		\theta_{\mathrm{weld}}^{m}
		-
		\theta_{\mathrm{target}}
		\right)
		\frac{I^{m}-I^{m-1}}
		{\theta_{\mathrm{weld}}^{m}-\theta_{\mathrm{weld}}^{m-1}}.
	\end{equation}
	
\begin{figure}[!htbp]
		\centering
		\begin{tikzpicture}[
			node distance=1.5cm,
			auto,
			block/.style={rectangle, draw, text width=7.5cm, align=center, rounded corners, minimum height=1cm, font=\small},
			innerblock/.style={rectangle, draw, text width=7cm, align=center, minimum height=0.8cm, font=\small},
			decision/.style={diamond, draw, text width=3.0cm, align=center, minimum height=1cm, font=\small, aspect=1.3},
			cloud/.style={draw, ellipse, align=center, minimum height=0.8cm, font=\small},
			line/.style={draw, -{Stealth[scale=1.2]}, thick}
			]
			
			\node [cloud] (start) {Start Resolution};

			\node [block, below=0.5cm of start] (init)
			{\textbf{Initialization}\\
				Choose initial current $I^{0}$, initialize the temperature field $\theta_h$, set $m=0$};
			
			\node [block, below=0.5cm of init] (outer)
			{\textbf{OUTER LOOP: Root-Finding Procedure}\\
				Set the excitation current: $I = I^{m}$};
			
			\node [block, below=0.5cm of outer] (middle)
			{\textbf{MIDDLE LOOP: Multiphysics Coupling}\\
				Update temperature-dependent material properties
				$\rho(\theta_h)$ and $\mu_{\mathrm{eff}}(\theta_h)$};
			
			\node [innerblock, below=0.5cm of middle] (em)
			{\textbf{INNER LOOP 1: Electromagnetic Solver}\\
				Solve for $\underline{\mathbf{h}}_h$\\
				Fixed-point iteration for
				$\mu_{\mathrm{eff}}(|\underline{\mathbf{h}}_h|)$};
						
			\node [innerblock, below=0.5cm of em] (source)
			{\textbf{Source Evaluation}\\
				Compute time-averaged Joule losses
				\[
				Q_{\mathrm{src}}
				=
				\frac12
				\rho(\theta_h)
				\left|
				\mathbf{curl}\,
				\underline{\mathbf{h}}_h
				\right|^2
				\]};
			
			\node [innerblock, below=0.5cm of source] (thermal)
			{\textbf{INNER LOOP 2: Thermal Transport Solver}\\
				Solve for
				$\theta_h$\\
				Newton--Raphson method};
			
			\node [decision, below=0.5cm of thermal] (check_mid) {$\left\|
				\theta_h-\theta_h^{\mathrm{prev}}
				\right\|
				\le
				\epsilon_{\mathrm{coupling}}$?};
			
			\node [decision, below=0.5cm of check_mid] (check_out)
			{$
				\left|
				\theta_{\mathrm{weld}}^{m}
				-
				\theta_{\mathrm{target}}
				\right|
				\le
				\epsilon_{\mathrm{target}}
				$?};
			
			\node [block, left=0.7cm of check_out, text width=5.6cm] (secant)
			{\textbf{Modified Secant Update}\\
				\textbf{If $m=0$:}
				$ I^{m+1}=I^{m}
				\sqrt{\dfrac{\theta_{\mathrm{target}}}
					{\theta_{\mathrm{weld}}^{m}}} $\\[1mm]
				\textbf{If $m\ge1$:}
				standard secant update\\
				Increment index:
				$m = m+1$};
			
			\node [block, below=0.5cm of check_out] (post) {\textbf{Post-Processing Operations}};
			\node [cloud, below=0.5cm of post] (end) {End Resolution};
			
			\path [line] (start) -- (init);
			\path [line] (init) -- (outer);
			\path [line] (outer) -- (middle);
			\path [line] (middle) -- (em);
			\path [line] (em) -- (source);
			\path [line] (source) -- (thermal);
			\path [line] (thermal) -- (check_mid);
			
			\path [line] (check_mid.east) -- node [near start, above] {No} ++(2.5cm,0) |- (middle.east);
			\path [line] (check_mid) -- node [left] {Yes} (check_out);
			
			\path [line] (check_out) -- node [above] {No} (secant);
			\path [line] (secant.north) |- (outer.west);
			
			\path [line] (check_out) -- node [left] {Yes} (post);
			\path [line] (post) -- (end);
			
		\end{tikzpicture}
		\caption{Resolution strategy flowchart}
		\label{fig:coupling_flowchart}
	\end{figure}
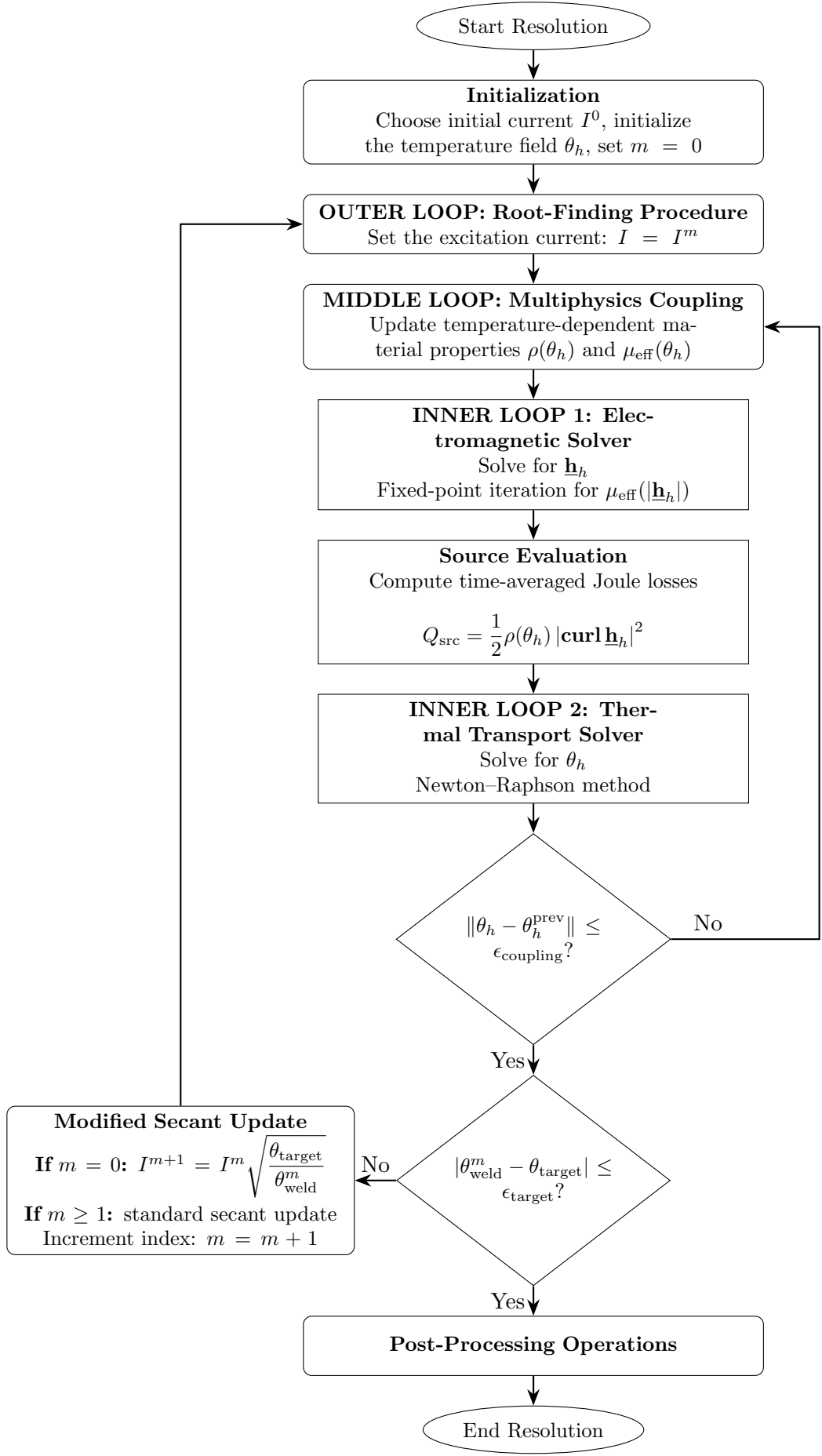

	For each prescribed current value $I^{m}$, a Picard fixed-point iteration resolves the coupling between the electromagnetic and thermal formulations. Within each outer Picard iteration, the coupled problem is decomposed into two specialized nonlinear subproblems:
	\begin{enumerate}
		\item \textbf{Electromagnetic solver:} The time-harmonic formulation~\eqref{eq:weak_separated} is solved using the current estimate of the temperature-dependent material properties. When nonlinear magnetic materials are present in the tube or impeder domains, an inner fixed-point iteration updates the effective permeability $\mu_{\mathrm{eff}}(|\underline{\mathbf{h}}|)$ until the magnetic residual satisfies the prescribed convergence criterion (inner loop~1 in Fig.~\ref{fig:coupling_flowchart}).
		
		\item \textbf{Thermal transport solver:} The nonlinear heat transfer problem is solved using a Newton--Raphson algorithm implemented through GetDP's \texttt{IterativeLoop} and \texttt{GenerateJac} functions (inner loop~2 in Fig.~\ref{fig:coupling_flowchart}).
	\end{enumerate}
	
	Once all nested iterations satisfy their respective convergence criteria, the converged electromagnetic and thermal fields are post-processed to extract secondary quantities, including coil voltage according to \eqref{eq:voltage}.
	
	\section{Geometry and Mesh Generation}

	\subsection{Tube Parametrization and Velocity Field}

\begin{figure} [!h]
		\centering
		\includegraphics[width=0.6\linewidth]{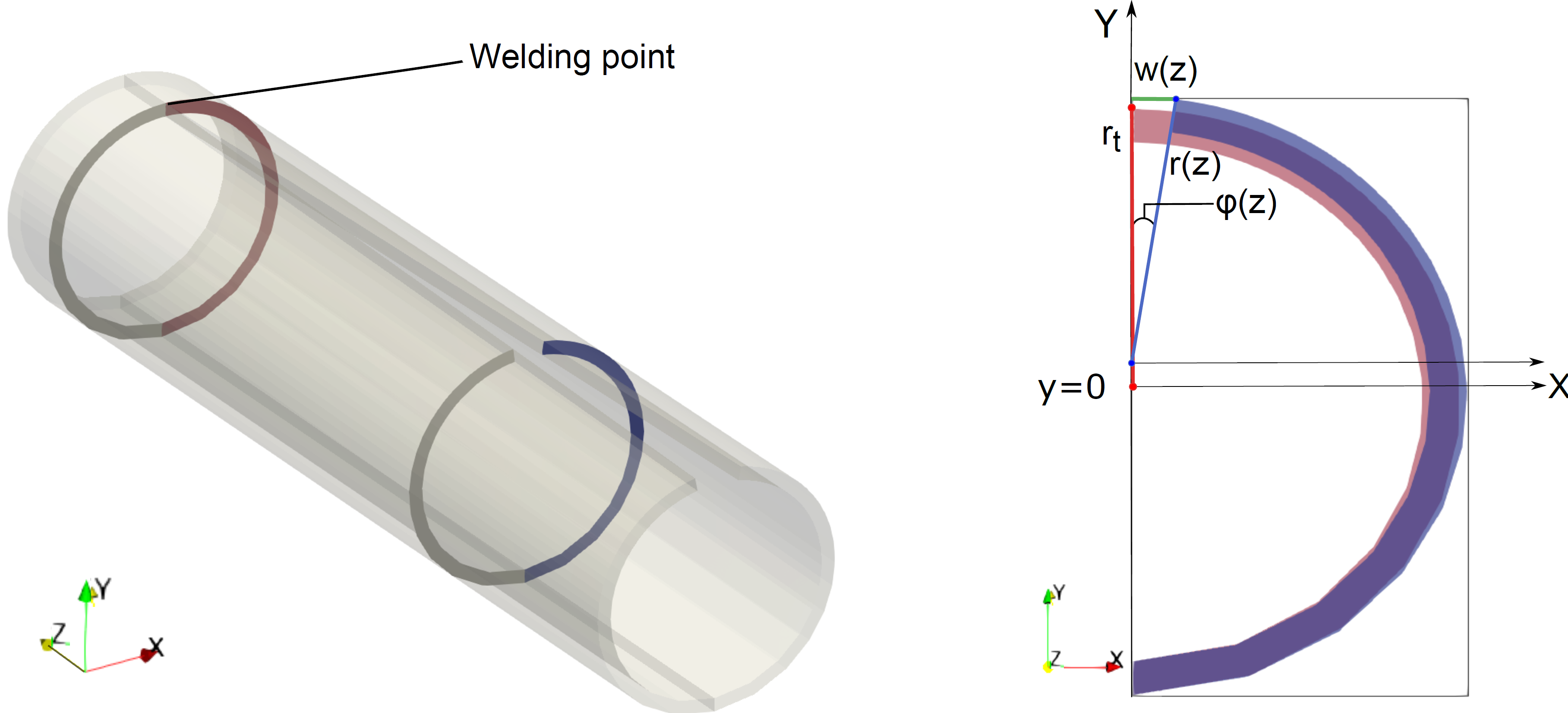}
		\caption{Parametrization of the strip cross-section}
		\label{fig:pda_geometry}
	\end{figure}

	To model the advective heat transport within the tube, the underlying geometric parametrization of the strip cross-section must be established first. Based entirely on this, the non-uniform advection velocity field $\mathbf{v}$ can subsequently be derived. As the material moves along the longitudinal axis $z$ at a nominal line speed $v_{\text{feed}}$, the continuous transition of the geometric domain defines the kinematic streamlines, which ultimately govern the velocity-driven transport of the thermal field before it reaches the weld apex ($z = 0$, see Fig. \ref{fig:pda_geometry}).
	
	To define the parametrized geometry, a slightly modified version of the approach presented in \cite{asperheim2024numerical} is adopted. We introduce a reference mapping
	\[
	X:(\alpha,\rho;r,\phi)\mapsto(x,y),
	\]
	where the geometric parameters $r$ and $\phi$ are functions of the local opening width $w$, see Fig.~\ref{fig:pda_geometry}. The parameter $\rho\in[0,1]$ denotes the normalized coordinate through the sheet thickness $h$, while $\alpha$ is the tangential angular coordinate. When welded ($w=0$), the local radius and opening angle satisfy $r=r_t$ and $\phi=0$ (see Fig.~\ref{fig:pda_geometry}), and the mapping is given by
	
	\begin{equation}
		X(\alpha, \rho; r_t, 0) = 
		\begin{pmatrix}
			\left(r_t + (\rho - 1)h\right) \sin \alpha \\[4pt]
			\left(r_t + (\rho - 1)h\right) \cos \alpha
		\end{pmatrix}
		\quad \text{for} \quad \alpha \in [0, \pi], \; \rho \in [0, 1]
	\end{equation}
		
	For an open cross-section profile where the opening width $w(z) > 0$ (see Fig.~\ref{fig:pda_geometry}), the sheet is modeled as an open circular arc characterized by a local outer radius $r(w)$ and a shifting opening angle $\phi(w)$. Invoking the physical constraint that the length of the outer diameter (OD) arc is strictly preserved during deformation, the local geometric variables must satisfy the algebraic system:
	
	\begin{align}
		(\pi - \phi) r &= \pi r_t \\
		w &= r \sin \phi
	\end{align}
	which is solved numerically for any given $w(z)$ to yield the geometric parameters $(r, \phi)$. The mapping for the deformed cross-section is therefore:
	
	\begin{equation}
		X\big(\alpha, \rho; r(w), \phi(w)\big) = 
		\begin{pmatrix}
			\left(r(w) + (\rho - 1)h\right) \sin \alpha \\[4pt]
			\left(r(w) + (\rho - 1)h\right) \cos \alpha + r(w) - r_t
		\end{pmatrix}
		\quad \text{for} \quad \alpha \in [\phi(w), \pi], \; \rho \in [0, 1]
	\end{equation}
	
	To evaluate the velocity components at an arbitrary spatial coordinate $(x, y, z)$ on the discretized finite element mesh, the local material parameters $(\alpha, \rho)$ must be reconstructed. The analytical inverse mapping $X^{-1}\big(x, y; w\big) = (\alpha, \rho)$ is derived as:
	\begin{align}
		\alpha &= \operatorname{atan2}\left(x, \, y - r(w) + r_t\right) \\[6pt]
		\rho &= \frac{\sqrt{x^2 + \left(y - r(w) + r_t\right)^2} - r(w)}{h} + 1
	\end{align}
		
	Assuming steady-state material flow where the material parameters $(\alpha, \rho)$ remain constant along a given streamline, the velocity field $\mathbf{v} = (v_x, v_y, v_z)^T$ across the forming domain is coupled to the line feed speed via the chain rule. Using the parameters retrieved from the inverse mapping, the complete 3D kinematic velocity vector is computed as:
	\begin{equation}
		\mathbf{v}(x,y,z) = 
		\begin{pmatrix}
			\dfrac{\partial x}{\partial w}(\alpha, \rho; w) \dfrac{\partial w}{\partial z} v_{\text{feed}} \\[10pt]
			\dfrac{\partial y}{\partial w}(\alpha, \rho; w) \dfrac{\partial w}{\partial z} v_{\text{feed}} \\[10pt]
			v_{\text{feed}}
		\end{pmatrix}
	\end{equation}
	where the shape sensitivities $\partial x / \partial w$ and $\partial y / \partial w$ are evaluated either analytically, as in \cite{asperheim2024numerical}, or via finite differences using a controlled perturbation $dw$.
	
	\subsection{Parameterized Geometric Modeling in Gmsh}
	\label{ssec:geom_gmsh}
	
	The three-dimensional geometry and the corresponding finite element mesh are generated using the open-source mesh generator Gmsh \cite{geuzaine2009gmsh}. Gmsh provides a script-based environment for parametric geometry construction through the Open CASCADE Technology (OCCT) CAD kernel \cite{opencascade} and is tightly integrated with the GetDP solver. By exploiting its support for variables, logical expressions, loops, and conditional statements, a fully parameterized model of the high-frequency induction tube welding system is constructed.
	
\begin{figure} [!h]
		\centering
		\includegraphics[width=0.8\linewidth]{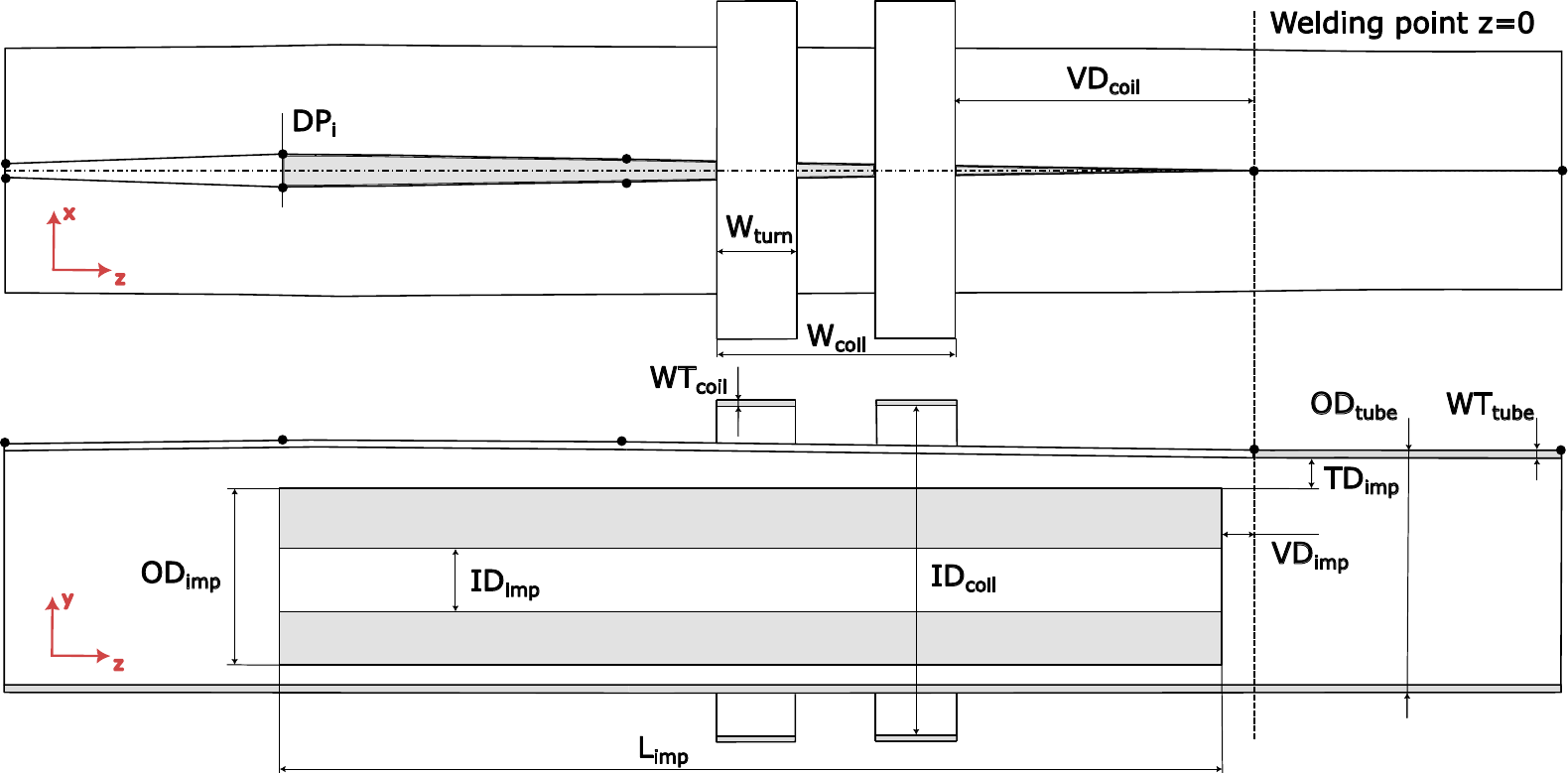}
		\caption{Parametrization of the tube welding system}
		\label{fig:geom_param}
	\end{figure}

	Within this framework, a unified set of global parameters completely defines the geometry of the multi-domain computational model. These parameters describe the dimensions of the welded tube $\Omega_t$, the induction coil turns $\Omega_c^i$, and the magnetic impeder $\Omega_{imp}$, together with their relative spatial positions (Fig.~\ref{fig:geom_param}). An additional set of parameters controls the mesh quality throughout the computational domains.
	
\begin{figure} [!h]
		\centering
		\includegraphics[width=0.8\linewidth]{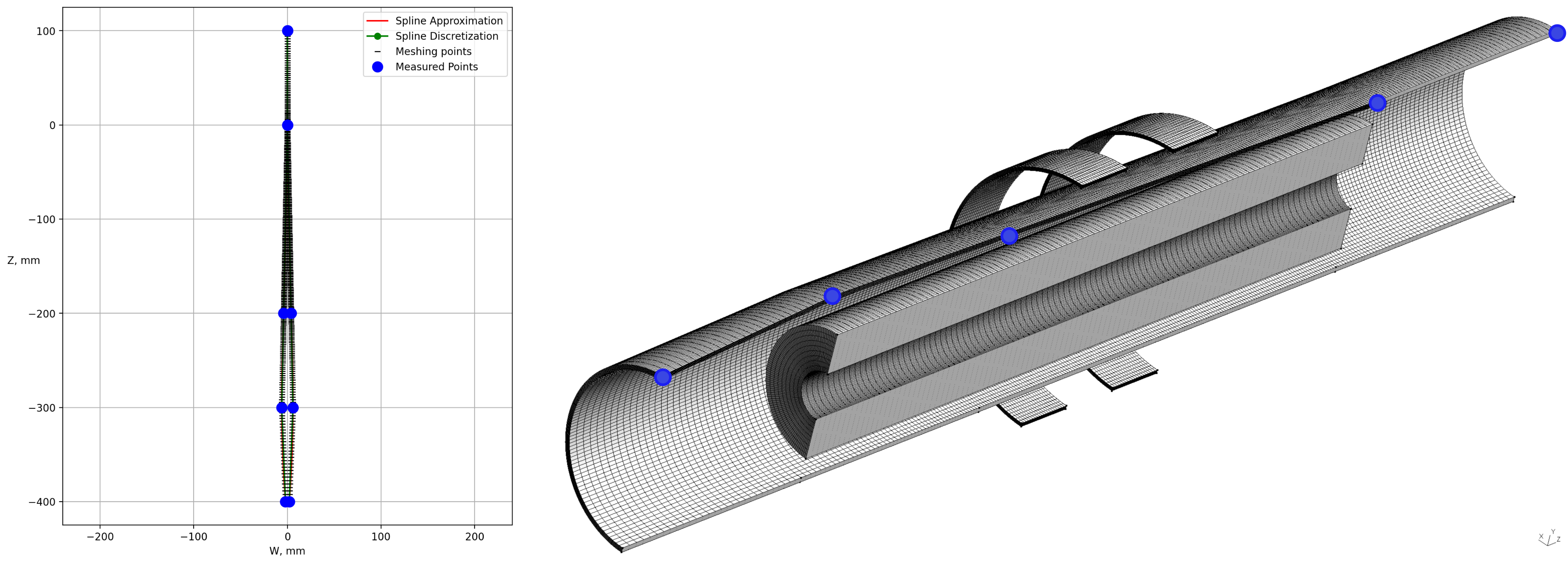}
		\caption{Approach to geometry and mesh modeling}
		\label{fig:mesh}
	\end{figure}

	The reconstruction of the strip geometry requires establishing the open profile configuration of the tube. Because of the shape complexity, a practical approach is to define it piece-wise by discretizing the physical profile along the longitudinal axis to the required accuracy. To achieve this, a standalone software tool was developed to execute the following geometric workflow (see fig. \ref{fig:mesh}):
	
	\begin{itemize}
		\item \textbf{Control points definition:} Manual placement of control points representing the outer edge of the tube strip. 
		\item \textbf{Continuous spline interpolation:} Automatic construction of a spline passing through the control points to establish a smooth, continuous profile.
		\item \textbf{Piecewise linearization:} Semi-automatic approximation of the spline by a set of linear segments suitable for export to a Gmsh geometry generation script.
		\item \textbf{Algorithm-assisted one-dimensional mesh construction:} Generation of a one-dimensional mesh along the piecewise defined segments. The mesh density is specified interactively by the user to account for electrical current distributions and provide the optimal resolution where it is needed. Smooth transitions in element size along the edge are achieved using compatible geometric progression-based node distributions.
	\end{itemize}

	\section{Validation}
	\label{sec:validation}
	
	\subsection{Validation Case Specification}
	
	To evaluate the accuracy and robustness of the developed framework, a validation case was established using operational data from an existing tube welding line. The benchmark dataset originates from a Weldac high-frequency induction welding installation. The geometric and process parameters defining this reference case are compiled in Table~\ref{tab:validation_parameters}.
	
	The total longitudinal length of the computational domain is defined manually based on the assumption that the induced current density becomes negligible at an upstream distance of approximately six tube outer diameters ($\approx 6\,\text{OD}_{\text{tube}}$) and a downstream distance of approximately $500\,\text{mm}$ from the welding apex, as demonstrated in previous work \cite{asperheim2024numerical}.

	\begin{table}[htbp]
		\centering
		\caption{Geometric, electrical, and process parameters for the steel 304 validation case}
		\label{tab:validation_parameters}
		\begin{tabular}{l l c}
			\hline
			\textbf{Category} & \textbf{Parameter} & \textbf{Value} \\ \hline
			\textit{Tube Profile} & Material 		 & AISI304 \\
			& Outer Diameter (OD\textsubscript{tube}) & 31.75 mm \\
			& Wall Thickness (WT\textsubscript{tube}) & 1.5 mm \\
			& Anchor points (DP\textsubscript{i}=(D, z) mm) & (8.0, -400.0) \\
			& 									  & 8.0, -200.0) \\
			&									  & (6.75, -150.0) \\ 
			&									  & (5.0, -100.0) \\ 
			&									  & (2.75, -50.0) \\ 
			&									  & (0, 0)  - welding point  \\
			&									  & (0, 50.0)   \\ \hline
			\textit{Induction Coil} & Number of Turns (N\textsubscript{coil}) & 2 \\
			& Turn Width (WT\textsubscript{turn}) & 20 mm \\
			& Inner Diameter (ID\textsubscript{coil}) & 40 mm \\
			& Width (WT\textsubscript{coil}) & 43 mm \\
			& Coil to Welding Point Distance (VD\textsubscript{coil}) & 43 mm \\ \hline
			\textit{Impeder} & Material & FeNiZnV \\
			& Outer Diameter (OD\textsubscript{imp}) & 13 mm \\
			& Inner Diameter (ID\textsubscript{imp}) & 3 mm \\
			& Length (L\textsubscript{imp}) & 200 mm \\
			& Impeder to Welding Point distance (VD\textsubscript{imp})  & 0 mm \\
			& Impeder to Tube Wall Distance (TD\textsubscript{imp})  & 5 mm \\ \hline 
			\textit{Electrical parameters} & Operating Frequency (f) & 193 kHz \\
			& Induction Coil RMS Current (I\textsubscript{RMS}) & 1500 A \\
			& Total Power (P\textsubscript{total}) & approx. 60 kW \\  \hline
			\textit{Feeding} & Line Feed Speed (v\textsubscript{feed}) & 49 m/min \\ \hline
		\end{tabular}
	\end{table}
	
	\subsection{Material Properties}
	
	\subsubsection{Electromagnetic properties}
	\label{sssec:elmag_properties}
	To solve the problem formulated in \eqref{eq:weak_separated}, the electrical resistivity $\rho(\theta)$ and the effective magnetic permeability $\mu_{\text{eff}}(|\underline{\mathbf{h}}|, \theta)$ must be defined across the entire computational domain $\Omega$.
	
	The electromagnetic properties within the tube domain $\Omega_t$ are defined as follows:
	\begin{itemize}
		\item $\rho(\theta)$: The electrical resistivity of AISI304 steel across a wide temperature range is adopted from \cite{chu1977electrical} and provided in Table~\ref{tab:ss304_resampled_grid}.
		\item $\mu_{\text{eff}}(|\underline{\mathbf{h}}|, \theta)$: Because AISI304 is a non-magnetic austenitic steel, its relative magnetic permeability remains constant ($\mu_r \approx 1$). This significantly simplifies the numerical treatment of the electromagnetic material properties within the tube domain ($\Omega_t$), yielding: $\mu_{\text{eff}}(|\underline{\mathbf{h}}|, \theta) = \mu_0$, where $\mu_0$ is the magnetic permeability of vacuum.
	\end{itemize}
	
	For the induction coil turns domain $\Omega_c^i$, the material parameters are given as a constant electrical resistivity $\rho(\theta) = 1.68 \times 10^{-8}\,\Omega\cdot\text{m}$, assuming efficient coil water cooling that maintains a stable operating temperature near $20^\circ\text{C}$, and $\mu_{\text{eff}}(|\underline{\mathbf{h}}|, \theta) = \mu_0$.
	
	For the surrounding atmospheric air domain $\Omega_a \in \Omega_c^C$, the condition $\rho(\theta) \to \infty$ is built directly into the formulation \eqref{eq:weak_separated}, and $\mu_{\text{eff}}(|\underline{\mathbf{h}}|, \theta) = \mu_0$. 
	
	Similarly to the air domain, the impeder domain $\Omega_{\text{imd}}$ is defined as non-conducting ($\Omega_{\text{imd}} \in \Omega_c^C$), leaving only its effective magnetic permeability $\mu_{\text{eff}}(|\underline{\mathbf{h}}|, \theta)$ to be specified. In contrast to many other induction heating applications, where magnetic cores are used as part of the coil to guide nearly all of the magnetic flux generated by the coil, an impeder intercepts and guides only a portion of the flux. Consequently, the magnetic flux distribution inside the welded tube depends on the spatial distribution of the impeder permeability. Because the permeability is nonlinear, this distribution depends on the magnetic field itself, making the problem inherently nonlinear and requiring a nonlinear solution.
	
\begin{figure} [!h]
		\centering
		\includegraphics[width=0.8\linewidth]{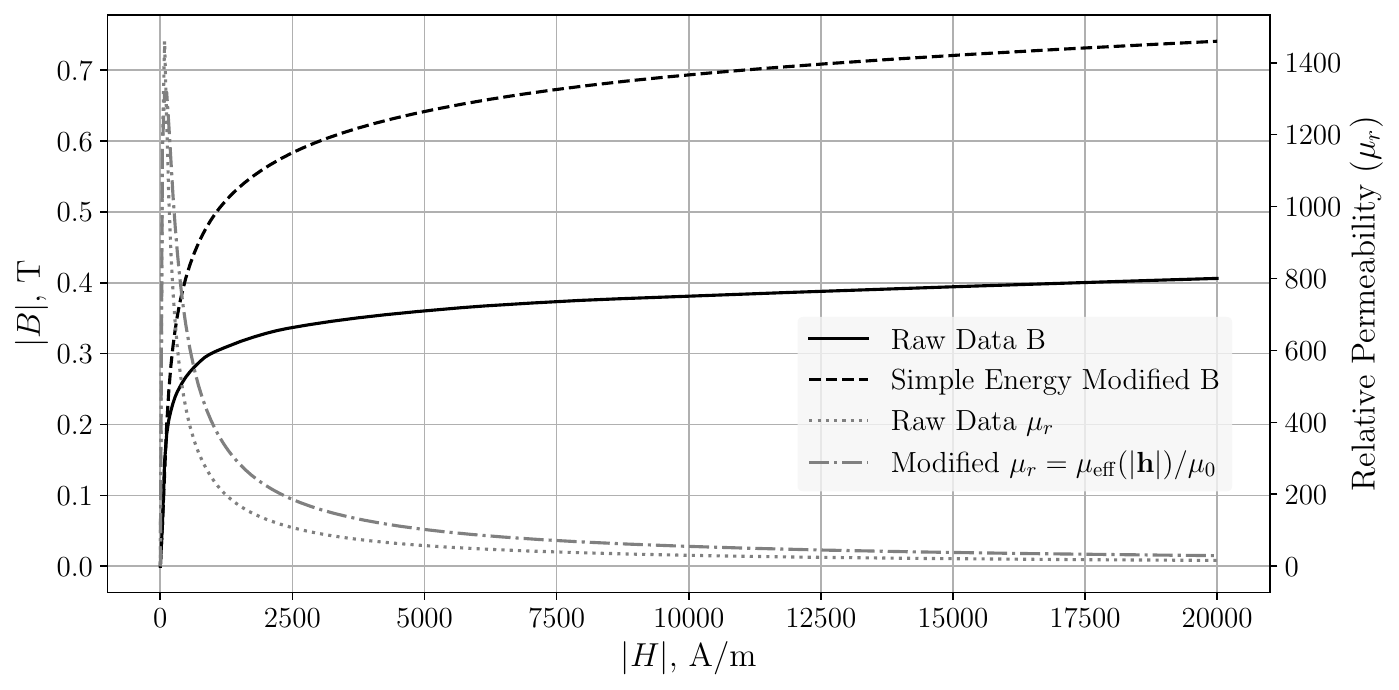}
		\caption{Magnetic properties of the FeNiZnV ferrite core}
		\label{fig:FeNiZnV_mag}
	\end{figure}

	Assuming efficient internal water cooling that maintains a stable operating temperature for the selected core material, it is sufficient to define the effective permeability solely as a function of the magnetic field intensity, $\mu_{\text{eff}}(|\underline{\mathbf{h}}|)$. In this work, the non-linear magnetic $B\text{-}H$ curve for the FeNiZnV ferrite was adopted from \cite{ASMMetalsHandbook1966} and subsequently mapped to $\mu_{\text{eff}}(|\underline{\mathbf{h}}|)$ via the co-energy transformation given in \eqref{eq:mu_effective_coenergy}. The resulting magnetic properties are illustrated in Fig.~\ref{fig:FeNiZnV_mag}.

	\subsubsection{Thermo-physical properties}

	For the heat transfer model \eqref{eq:thermal_discrete}, the transport behavior within the moving tube domain $\Omega_t$ is governed by the thermal properties of AISI304 steel. The temperature dependent density $\rho_m(\theta)$, specific heat capacity $c_p(\theta)$, and thermal conductivity $k(\theta)$ are adopted from \cite{chu1977electrical, kim1975thermophysical} and provided in Table~\ref{tab:ss304_resampled_grid}, alongside with the calculated thermal diffusivity $\lambda(\theta)$ required for SUPG implementation, according to \eqref{eq:peclet_definition}.
	
	\begin{table}[htbp]
	\centering
	\caption{Electro-thermo-physical properties of AISI304 steel}
	\label{tab:ss304_resampled_grid}
	\begin{tabular}{cccccc}
		\hline
		$\theta$ & $\rho(\theta)$ & $\rho_m(\theta)$ & $c_p(\theta)$ & $k(\theta)$ & $\lambda(\theta)$ \\
		&  \cite{chu1977electrical} & \cite{kim1975thermophysical} & \cite{kim1975thermophysical} & \cite{chu1977electrical} & (Calculated) \\
		($^\circ$C) & ($\Omega\cdot$m) & (kg/m$^3$) & (J/(kg$\cdot$K)) & (W/(m$\cdot$K)) & (m$^2$/s) \\ \hline
		0    & 6.961E-07 & 7894.0 & 510.03 & 14.40 & 3.577E-06 \\
		100  & 7.779E-07 & 7869.0 & 519.86 & 16.17 & 3.953E-06 \\
		200  & 8.494E-07 & 7832.9 & 533.13 & 17.83 & 4.269E-06 \\
		300  & 9.155E-07 & 7793.8 & 546.52 & 19.37 & 4.547E-06 \\
		400  & 9.762E-07 & 7752.9 & 560.21 & 20.82 & 4.794E-06 \\
		500  & 1.026E-06 & 7709.7 & 573.79 & 22.22 & 5.023E-06 \\
		600  & 1.077E-06 & 7664.1 & 587.20 & 23.61 & 5.246E-06 \\
		700  & 1.119E-06 & 7616.2 & 600.59 & 24.97 & 5.459E-06 \\
		800  & 1.152E-06 & 7565.7 & 613.98 & 26.24 & 5.649E-06 \\
		900  & 1.182E-06 & 7513.2 & 627.60 & 27.48 & 5.828E-06 \\
		1000 & 1.205E-06 & 7458.7 & 641.17 & 28.82 & 6.027E-06 \\
		1100 & 1.232E-06 & 7401.5 & 654.55 & 30.03 & 6.198E-06 \\
		1200 & 1.255E-06 & 7342.1 & 667.95 & 31.22 & 6.366E-06 \\
		1300 & 1.282E-06 & 7280.6 & 681.54 & 32.39 & 6.527E-06 \\
		1400 & 1.304E-06 & 7218.4 & 698.73 & 32.86 & 6.516E-06 \\
		1500 & 1.513E-06 & 6883.3 & 794.96 & 28.68 & 5.241E-06 \\
		1600 & 1.520E-06 & 6812.9 & 794.96 & 28.99 & 5.353E-06 \\
		1700 & 1.520E-06 & 6743.6 & 794.96 & 28.99 & 5.408E-06 \\ \hline
		\end{tabular}
	\end{table}

	\subsection{Approach to Meshing}
	
	The $h-\phi$ formulation presented in \eqref{eq:weak_separated} allows for a significant reduction in computational complexity and memory requirements compared to alternative 3D formulations of Maxwell's equations, such as the magnetic vector potential approach \cite{pellikka2013homology}. Nevertheless, full-scale 3D simulation of induction tube welding remains a challenging problem that requires careful mesh construction and optimization. To achieve this, the following factors must be considered:
	
	\begin{itemize}
		\item The mesh element size must be strictly related to the electromagnetic current penetration depth, which ranges between $0.8\,\text{mm}$ and $1.3\,\text{mm}$ for AISI304 steel within the process temperature range, and is approximately $0.15\,\text{mm}$ for copper at an operational frequency of $193\,\text{kHz}$. To accurately resolve the steep induced current gradients, a mesh density of at least three elements per penetration depth is typically sufficient when employing elements with linear basis functions.
		\item The current induced in the tube strip distributes across multiple paths and spans a large physical area \cite{asperheim2024numerical, jurgens1994three}. To accurately capture the volumetric Joule losses throughout the workpiece, the entire longitudinal length of the tube domain must be discretized with sufficient mesh quality.
		\item At the welding apex, the induced current abruptly changes direction and concentrates heavily, introducing a localized, extremely high current density. Consequently, the apex area must be sufficiently resolved.
		\item Air domain must be large enough to satisfy physically the condition \eqref{eq:bc_magnetic_wall}. In this work, the air domain is represented by a half-cylinder with diameter OD\textsubscript{tube}$\times 8$.
	\end{itemize}
	
	Edge Whitney elements are used to discretize of $\Omega_t$ and $\Omega_c^i \subset \Omega_c$ (see Sec.~\ref{sssec:elmag_discret} for details), making these conducting domains the primary contributors to the computational cost of the resulting system of linear equations. Fortunately, the topology of the tube domain $\Omega_t$, the induction coil turns  $\Omega_c^i$ and impeder $\Omega_{imp} $ allows for the use of a structured mesh, such as the one shown in Fig.~\ref{fig:mesh}. Implementing a structured mesh for the conducting domains provides several advantages, including rapid mesh generation, precise control over mesh density, element count, and element quality, hereby reducing the computational cost and memory requirements for the subsequent FEM solution. Furthermore, this approach can be readily implemented as part of the custom geometric parametrization developed within the Gmsh environment (see Sec.~\ref{ssec:geom_gmsh}). 
	
	In contrast, the air domain $\Omega_a$ should be discretized using an unstructured mesh, which allows for flexible adaptation of the element size and quality in regions where it is dictated by the physics of the simulated process, while enabling smooth interfacing with the structured meshes of the coil turns, impeder, and tube. 
	
	As shown in Sec.~\ref{sssec:cuts}, the chosen formulation requires the construction of topological cutting surfaces $\Sigma_c^i$ for each individual coil turn $\Omega_c^i$. Gmsh provides a homology solver plugin to automate this procedure \cite{pellikka2013homology}. The algorithmic approach is formulated as follows.
	
	For each individual induction coil turn $k \in \{1, \dots, N_c\}$, a complementary subdomain $\Omega_{\text{sub}}^k$ is established by gathering the remaining active turns:
	
	\begin{equation}
		\Omega_{\text{sub}}^k = \bigcup_{i=1, i \neq k}^{N_c} \Omega_c^i
	\end{equation}
	
	The topological space $\Omega_{\text{cut}}^k = \Omega_a \cup \Omega_t \cup \Omega_{\text{imp}} \cup \Omega_{\text{sub}}^k$ is passed to the homology solver for the calculation of the $k$-th independent thick cutting surface via the first cohomology group $H^1(\Omega_{\text{cut}}^k)$ \cite{pellikka2013homology, dlotko2010efficient}.

	\subsection{Simulation Results}
		
	The simulation was run skipping the outer loop in the resolution flowchart shown in Fig. \ref{fig:coupling_flowchart}. Since the current is already defined, the validation case targets the resulting temperature distribution.
	
	The primary reference metrics used for validation are the total power and the temperature at the welding apex. It should be noted that both benchmark values carry inherent engineering uncertainties:
	\begin{itemize}
		\item The total power is nominally specified at approximately 60~kW. This value is derived from the induction converter readings by subtracting estimated losses within the converter itself and the output matching circuit under operating conditions. The estimated accuracy is about $\pm 3\%$, yielding a reference confidence interval of 58--62~kW.
		\item Direct temperature measurement at the welding point is restricted and inherently inaccurate. However, standard industrial welding practice for achieving high-quality weld prescribes a temperature range of 1300--1400~$^\circ$C within the immediate weld zone in case of AISI304 steel. The target benchmark temperature is therefore established at $1350~^\circ\text{C} \pm 3.7\%$.
	\end{itemize}
		
	The simulation results are compiled in Table~\ref{tab:validation_results}. Both the total power $P_{\text{total}}$ and the temperature at the welding point $\theta_{\text{weld}}$ fall within the specified confidence intervals, validating the accuracy of the developed numerical framework. It should be noted that the current version of the model neglects magnetic core losses within the impeder. 
	
	\begin{table}[htbp]
		\centering
		\caption{Validation case results}
		\label{tab:validation_results}
		\begin{tabular}{ccccccc}
			\toprule
			$\underline{I}_{\text{RMS}}$ & $f$ & $\underline{V}_{\text{RMS}}$ & $P_{\text{tube}}$ & $P_{\text{coil}}$ & $P_{\text{total}}$ &  $\theta_{weld}$\\
			(A) & (kHz) & (V) & (W) & (W) & (W) & ($^\circ$C)\\
			\midrule
		
			 1500+0j & 193 & -39.93-219.42j & 57711.5 & 2185.3 &  59896.8 &  1325.73\\
			\bottomrule
		\end{tabular}
	\end{table}
	
\begin{figure} [!h]
		\centering
		\includegraphics[width=0.8\linewidth]{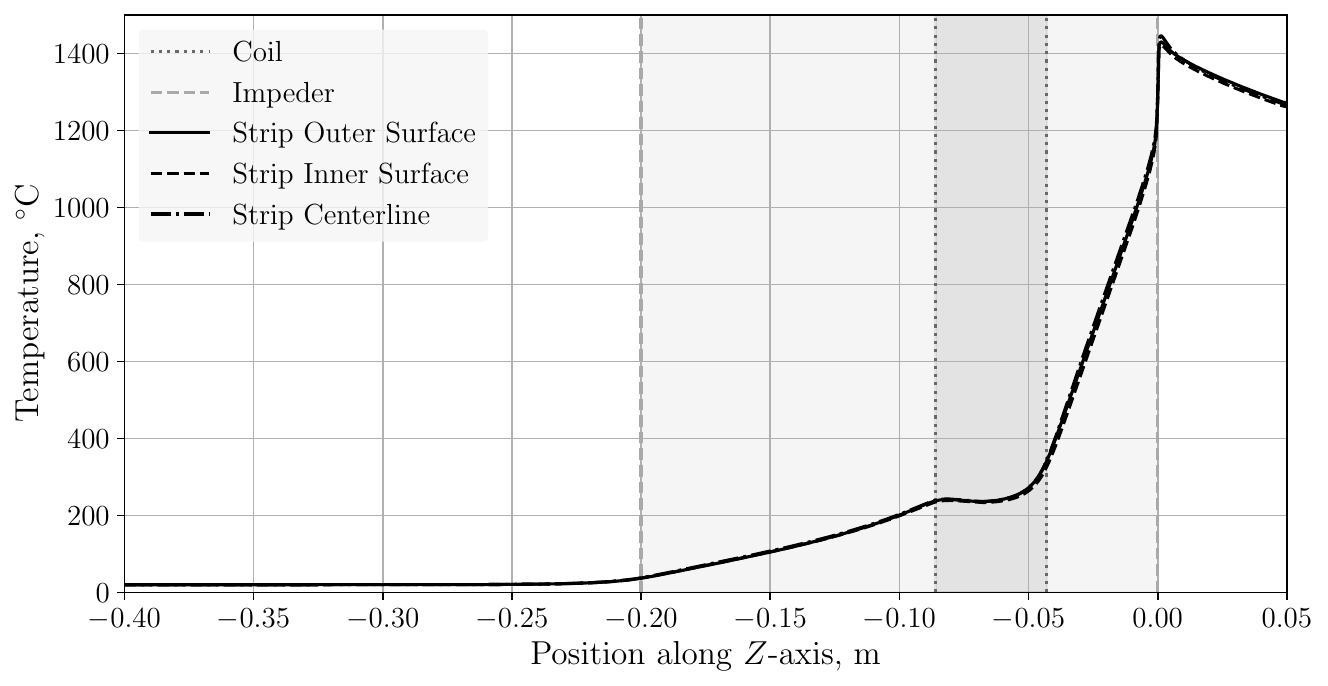}
		\caption{Temperature along the strip edge}
		\label{fig:T_along_edge}
	\end{figure}

	Post-processing capabilities of GetDP allow the automatic extraction of field quantities along predefined curves, which can be directly linked to the geometric parameterization (see Fig.~\ref{fig:geom_param}). One of the most informative and compact post-processing datasets for a process engineer is the temperature distribution along the strip edge, shown in Fig.~\ref{fig:T_along_edge}. The transparent gray regions indicate the locations of the induction coil and the magnetic impeder. A characteristic temperature drop is observed beneath the induction coil, reflecting the current and corresponding power density distribution along the edge \cite{asperheim2024numerical}. Note that the maximum temperature in the plot exceeds the target value at the welding point ($\theta_{weld}=1350 ^\circ C $), indicating a non-zero current density downstream of it. It is evident from the temperature distribution that the combination of the process parameters, geometry, and welded material results in a nearly uniform temperature through the strip thickness, as the corresponding temperature profiles are almost indistinguishable.
	
\begin{figure} [!h]
		\centering
		\includegraphics[width=0.8\linewidth]{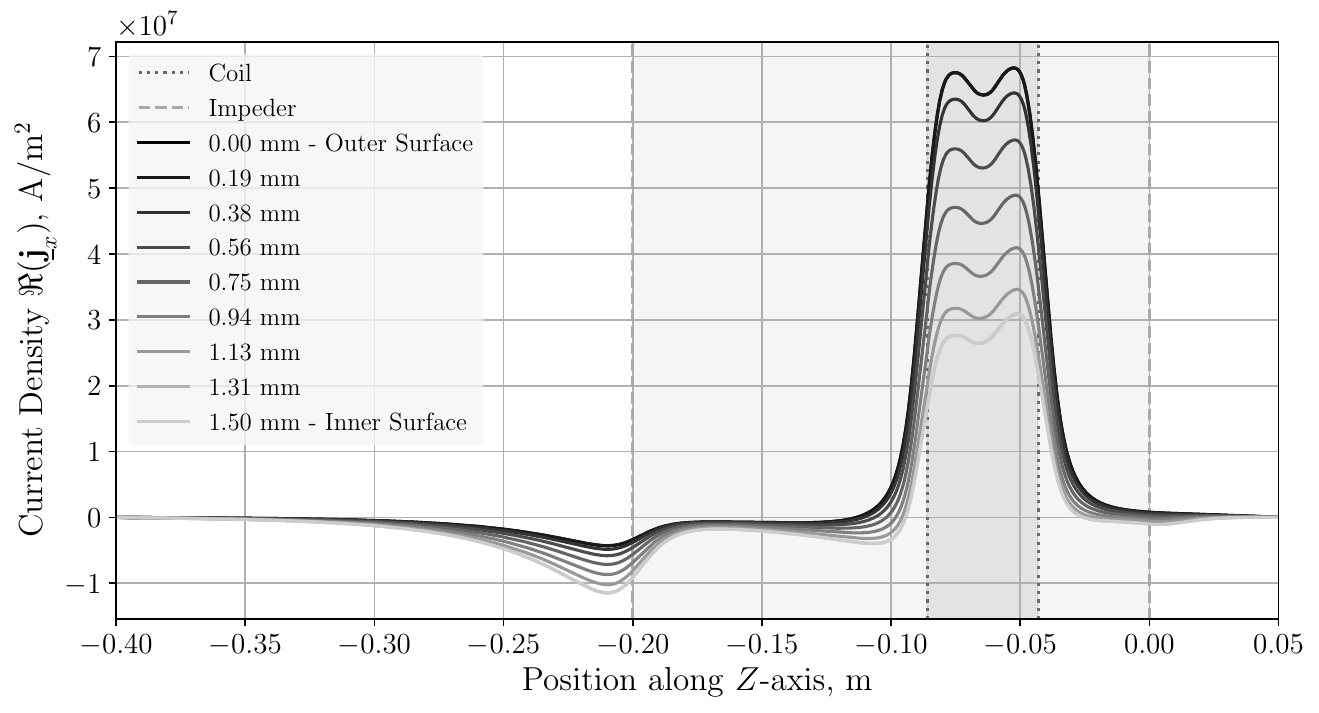}
		\caption{Current density $\Re(\underline{\mathbf{j}}_x)$ at the bottom of the tube (6 o'clock)}
		\label{fig:current_jx_re}
	\end{figure}

\begin{figure} [!h]
		\centering
		\includegraphics[width=0.8\linewidth]{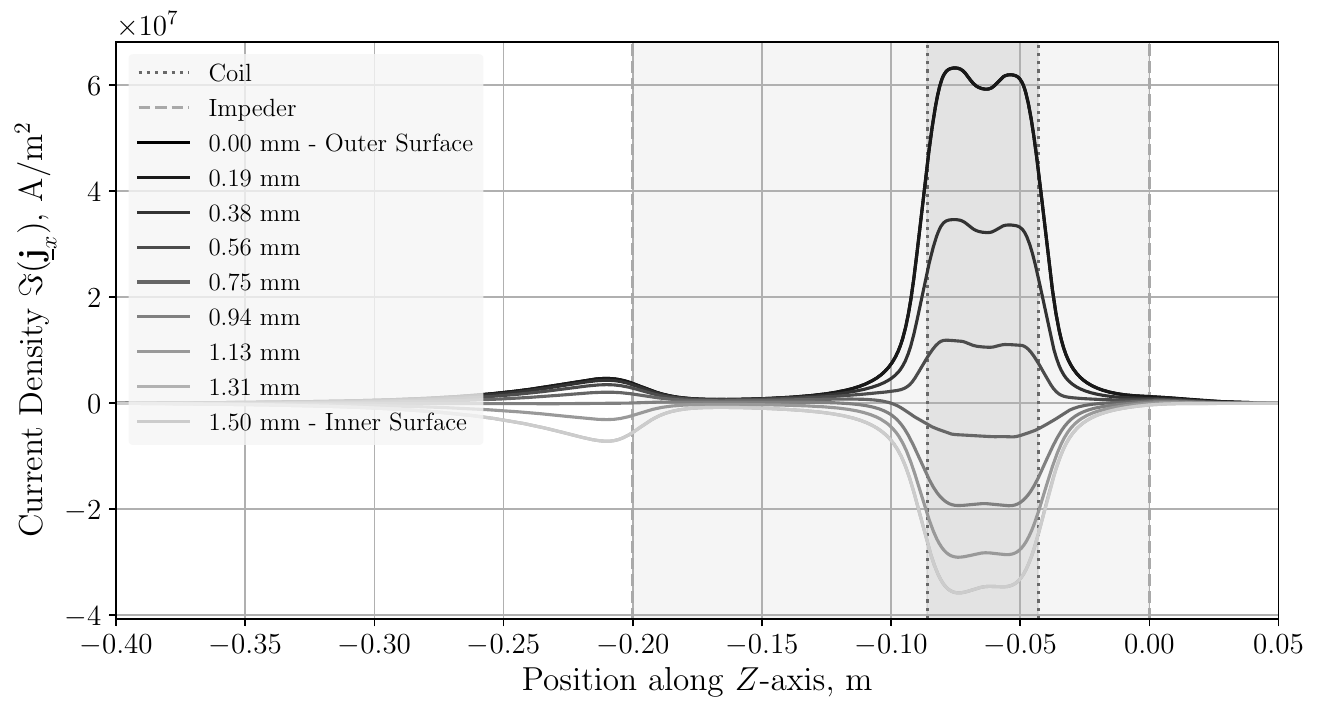}
		\caption{Current density $\Im(\underline{\mathbf{j}_x})$ at the bottom of the tube (6 o'clock)}
		\label{fig:current_jx_im}
	\end{figure}

	Another useful post-processing information reflecting the electromagnetic regime of the tube welding setup is the distribution of the $x$-component of complex current density $\underline{\mathbf{j}}_x = (\mathbf{curl} \, \underline{\mathbf{h}})_x$ at the bottom symmetry boundary of the tube strip cross-section ($\partial \Omega_{t, sym}$), see Fig. \ref{fig:current_jx_re} and \ref{fig:current_jx_im}. Due to the imposed symmetry boundary conditions \eqref{eq:bc_mirror}, the magnetic field $\underline{\mathbf{h}}$ at this location is purely tangential, forcing the current density $\underline{\mathbf{j}}$ to be purely normal to this plane ($x$-component only). Consequently, evaluating $\underline{\mathbf{j}}_x$ along this boundary directly captures the total current density. Furthermore, the real and imaginary parts of the current density $\underline{\mathbf{j}}_x$ contain information regarding both its magnitude and its phase shift relative to the coil current (which is defined as purely real). This detailed phase and magnitude information is crucial for analyzing induction welding systems, particularly when evaluating how the impeder’s position, dimensions, and material properties affect performance.

	\section{Investigation into the Effect of Impeder Magnetic Material Properties}
	
	\subsection{Motivation and design of the numerical experiment}
	
	The developed framework enables the analysis of a wide range of geometrical and electrical parameters governing the performance of a high-frequency induction tube welding system. One important and challenging aspects is the influence of the magnetic impeder. 
	
	The primary function of the impeder is to provide a low-reluctance path for the magnetic flux inside the tube, thereby minimizing the flux crossing the air gap between the impeder and the inner surface of the tube \cite{asperheim2024numerical}. As a result, the current induced on the inner surface (parasitic path) of the strip is significantly reduced. 
	
	It was already reasoned that the magnetic properties of impeders are of great importance (see Section \ref{sssec:elmag_properties}). Some published studies have considered soft magnetic composites (SMCs) as an alternative to ferrites for impeder materials \cite{muyskens2020improving, muyskens2023imat}, due to the distinctive magnetic properties of this type of material.
		
\begin{figure} [!h]
		\centering
		\includegraphics[width=0.8\linewidth]{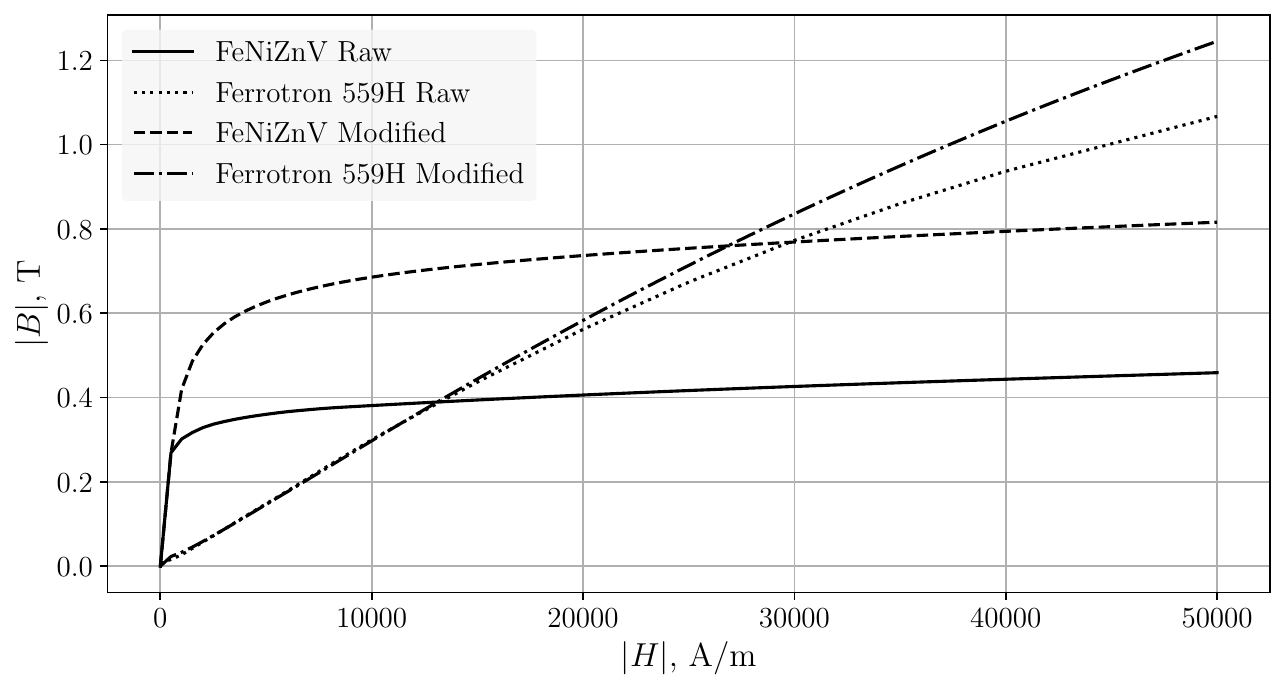}
		\caption{Magnetic properties of the FeNiZnV vs. Ferrotron 559H}
		\label{fig:FeNiZnV_vs_Ferrotron}
	\end{figure}

	Fig. \ref{fig:FeNiZnV_vs_Ferrotron} compares the original and modified B--H curves (calculated via the simple energy approach \eqref{eq:mu_effective_coenergy}) for the FeNiZnV ferrite and Ferrotron 559H SMC material \cite{fluxtrol_ferrotron_559h_original}.
 	It shows that there is a crossover point at a relatively high magnetic field strength (12–13 kA/m for the unmodified curves, corresponding to a magnetic flux density of approximately 0.4 T), at which the magnetic flux density (and thus the permeability) of Ferrotron 559H becomes higher. As a result, the reluctance of the impeder is reduced for a given cross-sectional area. To investigate the potential consequences of this effect, a numerical study was designed based on the following considerations:
	
	\begin{itemize}
		\item AISI304 was chosen as the tube material because it has already been shown to offer numerous computational advantages, including linear magnetic behavior of the steel tube and a relatively large current penetration depth. These properties make it possible to obtain accurate solutions within the developed framework, as demonstrated in Section~\ref{sec:validation}.
		
		\item A small tube diameter of 25.4~mm was selected to provide limited space for the magnetic impeder.
		
		\item A relatively large tube wall thickness of 3~mm was chosen to push the required current and power.
	\end{itemize}
	
	The design of the numerical experiment is provided in the Table \ref{tab:impeder_case}.

	\begin{table}[htbp]
		\centering
		\caption{Designed simulation study to compare FeNiZnV ferrite to Ferrotron 559H}
		\label{tab:impeder_case}
		\begin{tabular}{l l c}
			\hline
			\textbf{Category} & \textbf{Parameter} & \textbf{Value} \\ \hline
			\textit{Tube Profile} & Material 		 & AISI304 \\
			& Outer Diameter (OD\textsubscript{tube}) & 25.4 mm \\
			& Wall Thickness (WT\textsubscript{tube}) & 3.0 mm \\
			& Anchor points (DP\textsubscript{i}=(D, z) mm) & (5.28, -369.5) \\
			& 									  & (6.5, -122.0) \\
			&									  & (6.5, -72.0) \\ 
			&									  & (5.8, -59.5) \\ 
			&									  & (4.93, -47.0) \\ 
			&									  & (3.08, -23.5) \\ 
			&									  & (0.0, 0.0) - welding point  \\
			&									  & (0.0, 50.0)   \\ \hline
			\textit{Induction Coil} & Number of Turns (N\textsubscript{coil}) & 2 \\
			& Turn Width (WT\textsubscript{turn}) & 10 mm \\
			& Inner Diameter (ID\textsubscript{coil}) & 36 mm \\
			& Width (WT\textsubscript{coil}) & 25 mm \\
			& Coil to Welding Point Distance (VD\textsubscript{coil}) & 47 mm \\ \hline
			\textit{Impeder} & Material & \textbf{FeNiZnV vs.}\\
			&									& \textbf{Ferrotron 559H}\\
			& Outer Diameter (OD\textsubscript{imp}) & 13 mm \\
			& Inner Diameter (ID\textsubscript{imp}) & 5 mm \\
			& Length (L\textsubscript{imp}) & 200 mm \\
			& Impeder to Welding Point distance (VD\textsubscript{imp})  & -3 mm \\
			& Impeder to Tube Wall Distance (TD\textsubscript{imp})  & 3.2 mm \\ \hline 
			\textit{Electrical parameters} & Operating Frequency (f) & 200 kHz \\ \hline
			\textit{Thermal  parameters} & Weld temperature (T\textsubscript{weld}) & 1350 $^\circ\text{C}$ \\ \hline
			\textit{Feeding} & Line Feed Speed (v\textsubscript{feed}) & \textbf{40 m/min} \\ 
			&									&	\textbf{70 m/min} \\ 
			&									&	\textbf{100 m/min} \\ \hline
		\end{tabular}
	\end{table}
	
	\subsection{Simulation results}
		
	The simulation results for the two impeder materials are provided in Table \ref{tab:impeder_results}. At the lowest line feed speed of 40 m/min, the use of the FeNiZnV impeder results in a lower current and power consumption compared to the Ferrotron 559H, indicating higher permeability of the ferrite magnetic material at the required coil magnetomotive force (MMF). However, further increasing the line feed rate to 70 and 100 m/min changes this trend. At 100 m/min, using the Ferrotron 559H impeder reduces the required coil current from 5122.13 A to 3861.65 A (a 24.61\% reduction), while the power demand drops from approximately 314.12 kW to 231.25 kW (a 26.38\% reduction). The observed trend is expected and can be attributed to the magnetic properties of the materials shown in Fig. \ref{fig:FeNiZnV_vs_Ferrotron}.
	
		\begin{table}[htbp]
		\centering
		\caption{Performance comparison of the FeNiZnV and Ferrotron 559H impeders}
		\label{tab:impeder_results}
		\begin{tabular}{cccccc}
			\toprule
			 Impeder material & v\textsubscript{feed} & $|\underline{I}_{\text{RMS}}|$& $|\underline{V}_{\text{RMS}}|$  & $P_{\text{total}}$ & $L$ \\
			 				  & (m/min) & (A) & (V) & (W) & (nH)\\
			\midrule
			
			FeNiZnV 	  & 40 & 2372.28 & 379.6 & 94196.14 &  126.64 \\
			FeNiZnV       & 70 & 3819.2 & 557.8 & 194448.6 &  115.74 \\
			FeNiZnV 	  & 100 & 5122.13 & 712.82 & 314123 &  110.34 \\
			Ferroton 559H & 40 & 2477.65 & 389.23 & 97897.4 &  124.37 \\
			Ferroton 559H & 70 & 3231.43 & 505.32 & 164144.6 &  123.81 \\
			Ferroton 559H & 100 & 3861.65 & 601 & 231250.3 &  123.23 \\
			\bottomrule
		\end{tabular}
	\end{table}
	
	It is worth noting that the current version of the developed framework does not account for losses within the impeder, which are typically higher for soft magnetic composite (SMC) materials compared to classical ferrites. Although the magnitude of these losses is much lower than the power dissipated in the tube and coil, higher core losses will complicate the design of the impeder's water cooling system when SMCs are utilized. To address this limitation, the framework must be complemented by a model for impeder losses alongside a heat transfer model for the impeder domain $\Omega_{\text{imp}}$.
	
\begin{figure} [!h]
		\centering
		\includegraphics[width=0.8\linewidth]{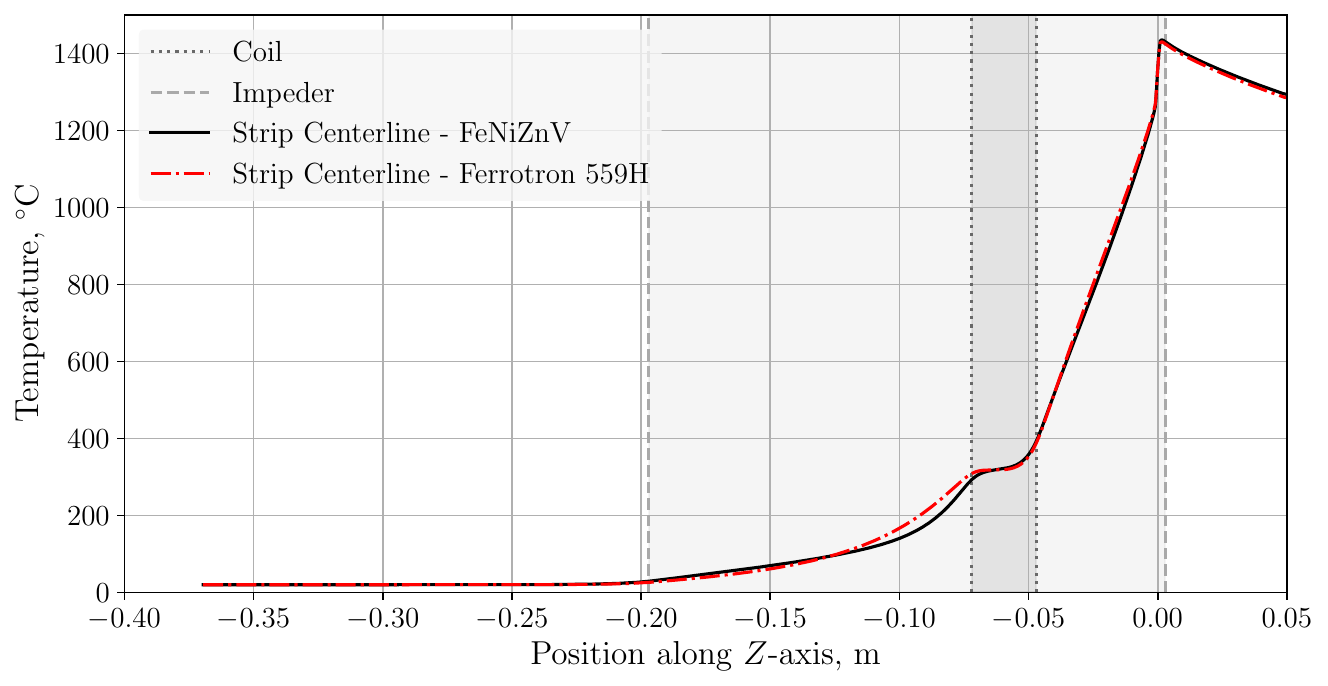}
		\caption{Temperature along the edge, v\textsubscript{feed} = 100 m/min: FeNiZnV vs. Ferrotron 559H impeders}
		\label{fig:T_along_edge_FeNiZnV_vs_Ferrotron_T}
	\end{figure}

	It is interesting to compare the performance of the two impeder materials under the most demanding operating conditions, namely at a feed rate of 100 m/min. As in Section~\ref{sec:validation}, we first consider the evolution of the temperature profile along the steel strip, this time evaluated only at the mid-thickness. The results are shown in Fig.~\ref{fig:T_along_edge_FeNiZnV_vs_Ferrotron_T}. Despite the large difference in current and power demand, the temperature profiles along the strip edge are almost indistinguishable. The current is determined automatically (see Fig. \ref{fig:coupling_flowchart}) to achieve the prescribed temperature at the welding point. The only slight difference between the temperature profiles obtained with the SMC and ferrite impeders can be observed upstream of the coil (to the left of it in the figure).
	
\begin{figure} [!h]
		\centering
		\includegraphics[width=0.8\linewidth]{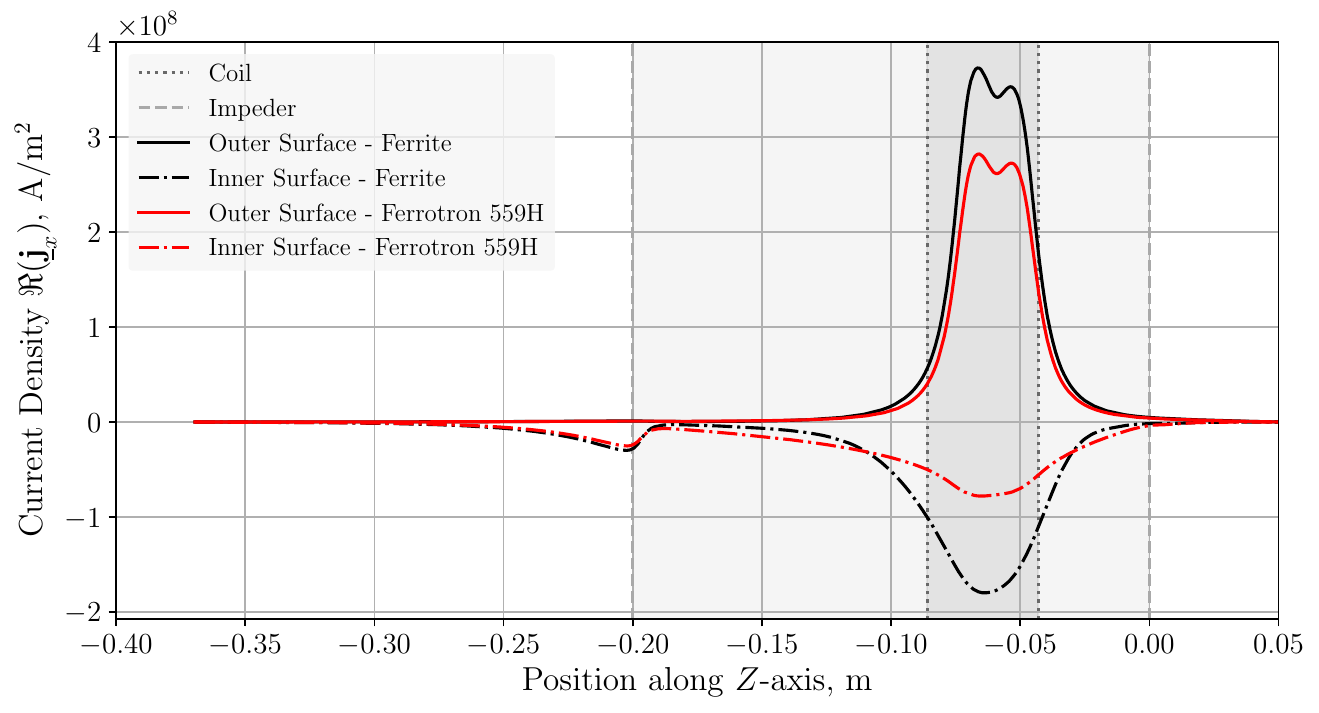}
		\caption{Current density $\Re(\underline{\mathbf{j}}_x)$ at the bottom of the tube (6 o'clock), v\textsubscript{feed} = 100 m/min: FeNiZnV vs. Ferrotron 559H impeders}
		\label{fig:current_jx_re_Ferrite-Ferrotron}
	\end{figure}

\begin{figure} [!h]
		\centering
		\includegraphics[width=0.8\linewidth]{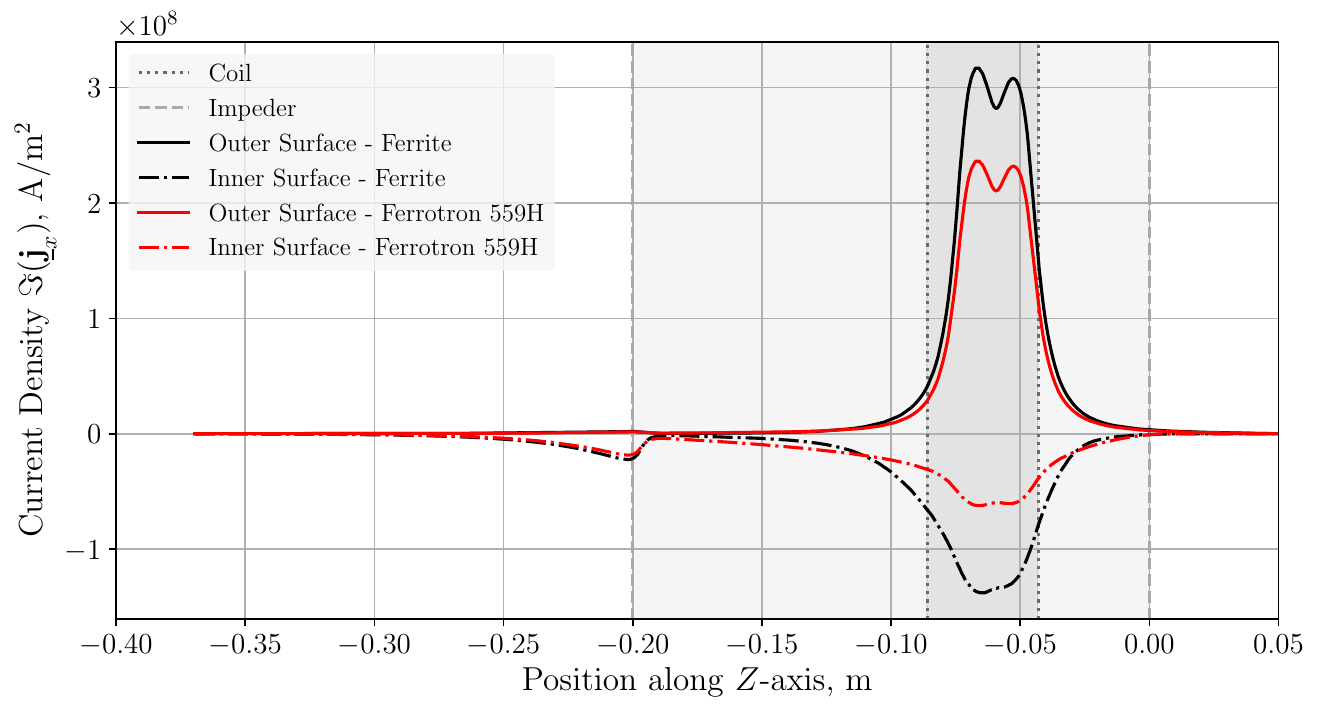}
		\caption{Current density $\Im(\underline{\mathbf{j}}_x)$ at the bottom of the tube (6 o'clock), v\textsubscript{feed} = 100 m/min: FeNiZnV vs. Ferrotron 559H impeders}
		\label{fig:current_jx_im_Ferrite-Ferrotron}
	\end{figure}

	 Clearly, the extra power generated when using the FeNiZnV impeder is dissipated along the circumference of the steel strip due to the higher magnitude of the outer and inner surface currents, resulting in a higher power density at these locations. Indeed, this effect can be clearly seen in Figs.~\ref{fig:current_jx_re_Ferrite-Ferrotron} and \ref{fig:current_jx_im_Ferrite-Ferrotron}, which show the distributions of the real and imaginary parts of the current density component $\underline{\mathbf{j}}_x$ at the bottom of the tube, as discussed in Section~\ref{sec:validation}. The current levels at both the outer and inner surfaces are significantly reduced when the Ferrotron 559H impeder is used. In addition, the lower inner-to-outer surface current ratio indicates the higher efficiency of the impeder when the SMC material is employed under conditions requiring a high MMF and where the available space for the impeder is limited.

	\section{Conclusions}

	In this work, a comprehensive 3D multiphysics finite element model has been presented for the simulation of high-frequency induction tube welding. By coupling the time-harmonic $h-\phi$ formulation with a stabilized quasi-stationary advection--diffusion thermal transport equation using the SUPG method, the model accurately captures the complex electromagnetic and thermal phenomena occurring at industrial line speeds. The framework was implemented using the open-source tools GetDP and Gmsh, with a minor modification to the GetDP kernel, enabling the dynamic evaluation of the element-wise grid Peclet number and the corresponding stabilization parameters.
	
	The accuracy and robustness of the proposed numerical model were validated against operational data from a commercial Weldac installation processing AISI 304 stainless steel tubes. 
	
	The applicability of the proposed framework was illustrated through a comparative numerical study investigating the influence of impeder magnetic material on the induction tube welding process. The performance of a conventional FeNiZnV ferrite impeder was compared with that of the soft magnetic composite (SMC) material Ferrotron 559H under demanding production conditions.	
	
	Given the complexity of high-frequency induction tube welding systems, several opportunities for further development and future research have been identified:
	
	\begin{itemize}
		\item \textbf{Influence of Welding Setup on Process Efficiency:} Conducting a systematic numerical investigation of how global system parameters, such as coil configuration, coil-to-apex distance, impeder position, forming setup and frequency influence the electrical efficiency and thermal performance of the welding process.

		\item \textbf{Impeder Losses and Heat Transfer:} Extending the multiphysics model to account for magnetic core losses together with convective fluid heat transfer within the impeder domain ($\Omega_{\text{imp}}$). This extension is essential for designing robust water-cooling systems for the impeders.
		
		\item \textbf{Advanced Impeder Geometries:} Developing a fully parameterized geometric framework capable of evaluating more advanced impeder topologies, including return-path impeders, slotted designs, and multi-material hybrid cores.
		
		\item \textbf{Shallow-Penetration Materials and Meshing:} Further optimization of the framework to simulate tube materials with very shallow electromagnetic penetration depths, such ferritic steels, aluminum, and copper alloys. Resolving these highly localized current layers requires a hybrid mesh incorporating structured hexahedral elements within the tube domain. 
		
		\item \textbf{Tailored Iterative Solvers:} Developing optimized, problem-specific iterative solvers and matrix preconditioning strategies to significantly reduce the computational time and memory requirements associated with large-scale industrial simulations involving millions of degrees of freedom.
	\end{itemize}

	\printbibliography

\end{document}